\documentclass[journal,twocolumn,oneside,letterpaper]{IEEEtran}

\usepackage{times}
\usepackage{cite}
\usepackage{mdwmath}
\usepackage[english]{babel}
\usepackage{epsfig}
\usepackage{easymat}
\usepackage{amsfonts}
\usepackage{bm}
\usepackage{amsbsy}
\usepackage{amsmath}
\usepackage{float}
\usepackage{cite}
\usepackage{graphicx}
\usepackage{balance}
\usepackage{subfigure}
\usepackage{latexsym}
\usepackage{amssymb}
\usepackage{txfonts}
\usepackage{wasysym}
\usepackage{mathbbol}
\usepackage{multirow}
\usepackage{stfloats}
\usepackage{enumerate}
\usepackage{flushend}
\usepackage{setspace}
\usepackage{color}
\usepackage{url}

\allowdisplaybreaks

\newtheorem{theorem}{Theorem}

\newtheorem{lemma}{Lemma}

\newtheorem{example}{Example}

\newcommand{\avr}{\textbf{E}}

\newcommand{\bt}[1]{\textbf{#1}}

\begin{document}

\title{On the Capacity of Multiple Access and Broadcast Fading Channels with Full Channel State Information at Low SNR
\author{ Zouheir Rezki~\IEEEmembership{Senior Member,~IEEE,}
 and Mohamed-Slim Alouini,~\IEEEmembership{Fellow,~IEEE,}}
\thanks{The authors are members of the KAUST Strategic Research Initiative (SRI) on Uncertainty Quantification in Science and Engineering.}
\thanks{Zouheir Rezki and Mohamed-Slim Alouini are with the Electrical Engineering Program, Computer, Electrical, and Mathematical Science and Engineering (CEMSE) Division, King Abdullah University of Science and Technology (KAUST), Thuwal, Makkah Province, Saudi Arabia. Email: \{zouheir.rezki,slim.alouini\}@kaust.edu.sa.}
\thanks{This work was funded by a Competitive Research Grant (CRG2) from the Office of Competitive Research Funding (OCRF) at KAUST.}
\thanks{Part of this work has been accepted for presentation in the 2013 IEEE International Conference on Communications (ICC'2013), Budapest, Hungary, and in the 2013 IEEE International Symposium on Information Theory (ISIT'2013), Istanbul, Turkey.}}

\maketitle

\begin{abstract}
We study the throughput capacity region of the Gaussian multi-access (MAC) fading channel with perfect channel state information (CSI) at the receiver and at the transmitters, at low power regime. We show that it has a multidimensional rectangle structure and thus is simply characterized by single user capacity points. More specifically, we show that at low power regime, the boundary surface of the capacity region shrinks to a single point corresponding to the sum rate maximizer and that the coordinates of this point coincide with single user capacity bounds. Inspired by this result, we propose an on-off scheme, compute its achievable rate, and show that this scheme achieves single user capacity bounds of the MAC channel for a wide class of fading channels at asymptotically low power regime. We argue that this class of fading encompasses all known wireless channels for which the capacity region of the MAC channel has even a simpler expression in terms of users' average power constraints only. Using the duality of Gaussian MAC and broadcast channels (BC), we deduce a simple characterization of the BC capacity region at low power regime and show that for a class of fading channels (including Rayleigh fading), time-sharing is asymptotically optimal.
\end{abstract}


\section{Introduction}\label{S1}
It is now widely accepted that energy efficiency is a key parameter in designing wireless communication systems. This has catalyzed interests of many researchers inside the information/communication theory communities in order to better understand performance limits of wireless communication in the low power regime, and develop new techniques to achieve/approach these limits, e.g., \cite{Verdu2002,Zheng2007,Li2010a,Lozano2003,Qiao2011}. For instance, in wide band communications, although the signal strength is generally very low, one can capitalize on the huge bandwidth and still achieve a high capacity \cite{Zheng2007,Raghavan2007a,Ray2007}. The low-SNR framework is also useful to model cellular networks in some specific cases \cite{Lozano2003,Beko2008a}, sensor networks where power saving is detrimental \cite{Chamberland2004,Jayaweera2006} and more generally any communication scenario where the bandwidth and the power are fixed, but the system degree of freedom is large enough such that the power per degree of freedom is very low \cite{Verdu2002,Porrat2007}. Mainly, there are two trends in the literature of communications at low-power regime. The first one focuses on studying fundamental limits in terms of channel capacity, error probability, error exponent, etc; e.g., \cite{Verdu2002,Zheng2007,Li2010a,Lozano2003,Qiao2011,Wu2007,Ray2007}. The second one deals more with signaling design and practical schemes that achieve these performance limits asymptotically, e.g., \cite{Perez-Cruz2010,Verdu2002,Gursoy2007,Gursoy2005,Wu2012}. A comprehensive list of references regarding communications at low-power regime and energy efficiency can be found in \cite{Porrat2007,Feng2012,Yang2004}  \
\begin{figure}[t]
  \centering
    \includegraphics[scale=0.27]{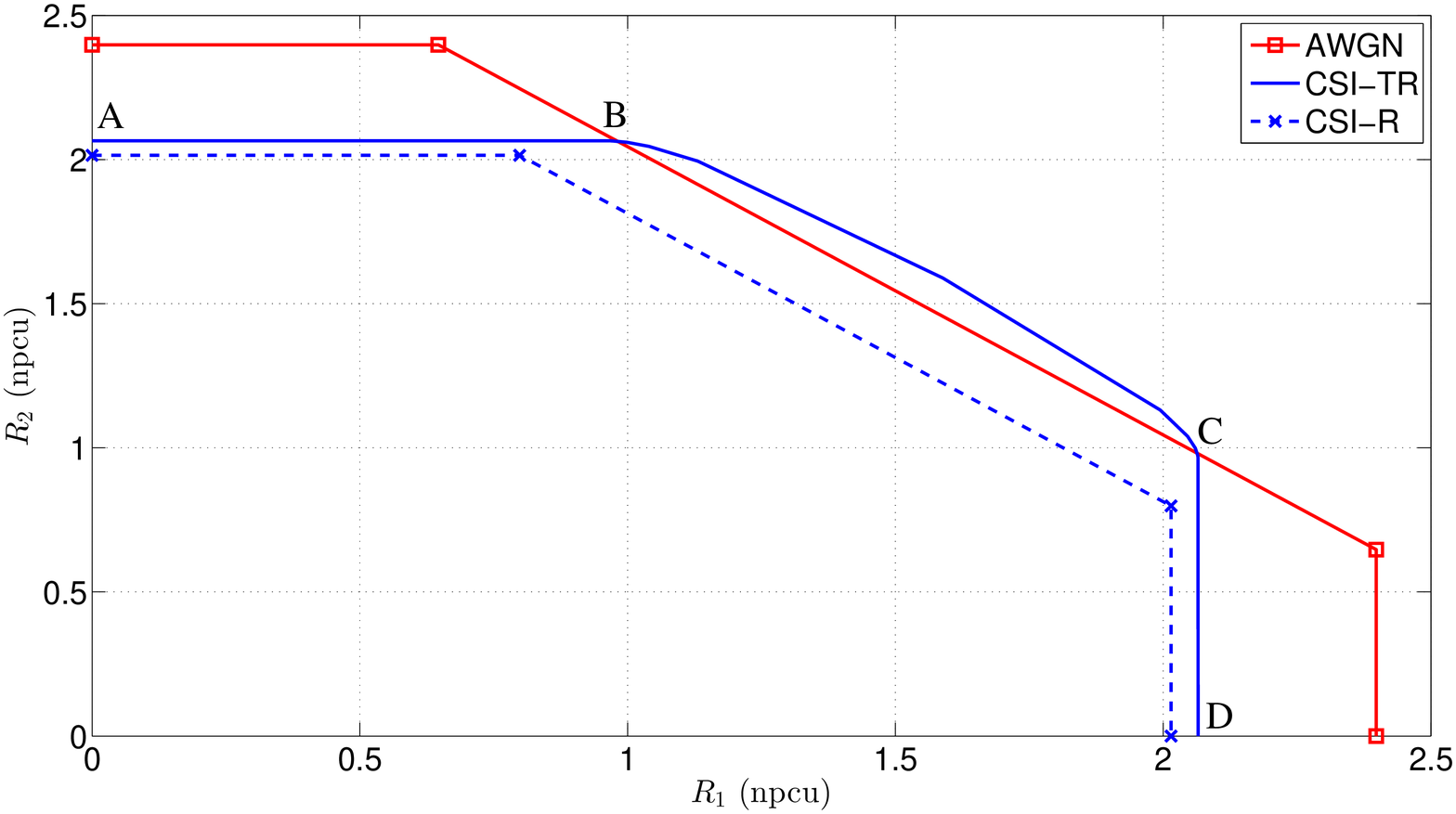}
        \caption{Capacity region of a 2-user MAC Rayleigh fading channel where $\bar{P}_1$ and $\bar{P}_2$ denote the average transmit power at the transmitters, respectively. The acronyms CSI-TR in the legend corresponds to the case where the CSI is available at both transmitters and at the receiver as well. The acronym CSI-R corresponds to the case where CSI is available at the receiver only.}
    \label{F1}
\end{figure}

In this paper, we aim at studying the throughput capacity region of fading multi-access channels (MAC) and fading broadcast channels (BC) with Gaussian noise, where perfect channel state information (CSI) at the receiver(s) (CSI-R) and at the transmitter(s) (CSI-T) is assumed, at low power regime. The throughput capacity of the Gaussian MAC fading channel has been derived in \cite{Tse1998}. Therein, it has been shown that each point on the boundary surface of the capacity region can be obtained by successive decoding and that the optimal rate and power allocations can be seen as the generalization of the single-user water-filling construction to MAC channels. The boundary surface is defined as the set of users' rates such that no component can be increased with other components remaining fixed, while staying in the capacity region \cite{Tse1998}. For instance, the capacity region of a 2-user symmetric MAC channel is depicted in Fig. \ref{F1}. The boundary surface is the curved line joining the points B and C. Differently from the nonfading Gaussian MAC channel where the boundary surface of the capacity region is a line with slope -1, there is no linear part on the boundary surface of the Gaussian MAC fading channels capacity region. Furthermore, since the horizontal line joining A and B and the vertical line joining C and D are completely defined by single user capacity bounds, then determining the boundary surface B-C is enough to fully characterize the MAC capacity region. However, to obtain each point on the boundary surface, a parameterized optimization problem must be solved.

On the other hand, the BC is generally a good model for downlink communications in cellular networks where a base station is sending common and/or independent messages to different users. The single-input single-output (SISO) Gaussian BC is by nature a degraded channel for which the capacity is known \cite{Cover2006a}. Even with fading along with perfect CSI at both the transmitter and the receivers (CSI-TR), the capacity region has been obtained in \cite{Li2001} and in the context of parallel Gaussian channels in \cite{Tse1997}. Here again, although the optimal power profile structure is essentially a water filling, an explicit characterization of the boundary capacity region seems difficult to obtain.

We focus in this paper on the low power regime (formally defined later) and analyze the capacity region of the Gaussian MAC fading channel. We show that interestingly, at low power regime, the MAC capacity region has a multidimensional rectangle structure, i.e. the boundary surface reduces to a single point. This point corresponds to the sum rate maximizer, the components of which are single-user capacity bounds. Then, we propose an on-off scheme and provide a necessary condition on the fading channels under which this scheme achieves the boundary point so that each user achieves single-user performance as if others were turned off. Using the duality of Gaussian MAC and BC, we provide a simple characterization of the BC capacity region at low power regime and show that for a class of fading channels (including Rayleigh fading), time-sharing is asymptotically optimal.

We note that the conference versions of this paper \cite{Rezki2013a,Rezki2013b} contain part of the results presented herein. However, we believe that the actual version is a substantial extension of the conference versions. We highlight below the main extra contributions beyond the conference versions:
\begin{itemize}
\item We present an explicit and simplified power profile (c.f. \eqref{E5A02} in Section \ref{S5A}) that achieves any point on the boundary of the Gaussian BC. Its simplicity is coming from the fact that the transmitter needs not solve multi-variable equations to set the optimal power. Instead, likewise in point-to-point communications, the optimal water-filling is derived by solving single-variable equations.
\item We provide a necessary and sufficient condition under which time-sharing is asymptotically optimal (c.f. \eqref{E5A04} in Section \ref{S5A}), thereby fully characterizing the class of fading channels for which time-sharing is asymptotically capacity-achieving. Illustrative examples (Example \ref{ex1} and Example \ref{ex2} in Section \ref{S5A}) are given to support this result.
\item We include the proof of Lemma \ref{L1} (Appendix \ref{Appendix1}) which is instrumental to the prof of our main result (Theorem \ref{T1}).
\item We insert the proof of Theorem \ref{T2} (Appendix \ref{Appendix3}) which we believe is an important result that characterizes the capacity of point-to-point communications at low power regime. While this result is instrumental to the optimality of the on-off power control for the MAC channel (c.f. \eqref{E403A} in Section \ref{S4}), it could be also of independent interest.
\item We reinforce our numerical results and include in all figures the capacity regions of the additive white Gaussian noise (AWGN) channel and the corresponding fading with channel state information at the receiver (CSI-R) only as benchmarks. We also include Fig. \ref{F10} as an evidence that time-sharing might not be optimal at low power regime for the log-logistic broadcast fading channel as discussed in Example \ref{ex2}.
\end{itemize}\

The remainder of this paper is organized as follows. Section \ref{S2} introduces our system model. Section \ref{S3} contains the main result of this paper along with its proof. An off-off scheme is proposed in Section \ref{S4} where the MAC capacity region is explicitly characterized for a class of fading channels. The capacity of BC is deduced in Section \ref{S5A}. In Section \ref{S5}, selected numerical results are provided. Finally, Section \ref{S6} concludes the paper.

\section{System Model}\label{S2}
We consider an uplink scenario where $K$ users are communicating with a base station. All terminals have a single antenna each, i.e., a SISO MAC channel. We focus on a discrete-time Gaussian MAC in which the received signal at the base station at time instant $n$, $n=1,\ldots,\infty$, is given by
\begin{equation}\label{E201}
  y(n) =\sum_{k=1}^{K} \, h_{k}(n) \, x_{k}(n) \, + \, v(n),
\end{equation}
where $h_{k}(n)$, $x_{k}(n)$ and $v(n)$ are complex random variables (r.v.) that represent the channel gain and the transmitted signal of user $k$ and the additive noise, respectively. For convenience, we denote a circularly symmetric complex Gaussian r.v., say $r$, with mean zero, and variance $\sigma^{2}$, as $r \sim \mathcal{CN}\left(0,\sigma^{2}\right)$. We assume without loss of generality a normalized Gaussian noise so that $v \sim \mathcal{CN}\left(0,1\right)$. Each user is constrained by an average transmit power $\bar{P}_k$.\

We also consider the $K$-user discrete-time Gaussian BC in which the received signal at user $k$, $k=1,\ldots,K$, is given by
\begin{equation}\label{E201B}
  y_{k}(n) = h_{k}(n) \, x(n) \, + \, v_{k}(n),
\end{equation}
where $x(n)$ and $v_{k}(n)$ are complex r.v. that represent the transmitted signal and the additive noise at user $k$. We also assume without loss of generality a normalized Gaussian noise so that $v_{k} \sim \mathcal{CN}\left(0,1\right)$. The transmitter is constrained by an average transmit power $\bar{P}$.\

We focus on fading processes with continuous probability density function (pdf) and with infinite support (although the later assumption is not mandatory). Furthermore, we assume that the channel gains of all users are independent, but not necessarily identically distributed. In addition, in both the MAC and the BC described by \eqref{E201} and \eqref{E201B}, respectively, perfect CSI-TR is assumed, implying that at time instant $n$, each terminal knows perfectly all channel gains $h_{k}(n)$, $k=1,\ldots,K$. For convenience, we let $\gamma_k=|h_k|^2$.
  We focus on asymptotically low power regime meaning that $\underset{k=1,\ldots,K}{\max} \bar{P}_k \rightarrow 0$ for the MAC and that $\bar{P} \rightarrow 0$ for the BC; and we say that $f(x) \stackrel{a}{\approx} g(x)$ if and only if $\underset{x \rightarrow a}{\lim}{ \; \frac{f\left(x\right)}{g\left(x\right)}}=1$. Inequalities $\tilde{<}$ and $\tilde{>}$ are defined analogously. The last definition extends to functions of severable variables where the definition of limits is standard. When it is clear from the context, we omit $a$ in $\stackrel{a}{\approx}$ for convenience. Unless the base is specified, all the $\log(\cdot)$ functions in this paper represent the natural logarithm functions. Finally, bold face letters indicate vectors of dimension $K$, i.e., $\bt{R}=\left(R_1,\ldots,R_K\right)$.

\section{MAC Capacity Region At Low Power Regime}\label{S3}
In this section, we first present our main result in Theorem \ref{T1} followed by the proof.
\begin{theorem}\label{T1}
Let $\mathcal{R}\left(\bar{\bt{P}}\right)$ be the multidimensional rectangle defined by the single user capacities, i.e., $\mathcal{R}\left(\bar{\bt{P}}\right)=\left\{\bt{R}: \, R_k \; \leq \; C_{k}\left(\bar{P}_{k}\right), \quad k=1,\ldots,K\right\}$, where $C_{k}\left(\bar{P}_{k}\right)$ is the single user capacity of user $k$, with average transmit power $\bar{P}_{k}$, and with perfect CSI-TR. For the SISO MAC described by (\ref{E201}), the capacity region $\mathcal{C}_{MAC}\left(\bar{\bt{P}}\right)$ coincides with $\mathcal{R}\left(\bar{\bt{P}}\right)$ at asymptotically low power regime. That is, for any point $\bt{R} \in $ on the boundary of the capacity region $\mathcal{C}_{MAC}\left(\bar{\bt{P}}\right)$, we have:
\begin{equation}\label{E301}
\underset{\bar{\bt{P}} \rightarrow \bt{0}}{\lim} \; \frac{R_k\left(\bar{\bt{P}}\right)}{C_{k}\left(\bar{\bt{P}}\right)}=1,
\end{equation}
for all $k=1,\ldots,K$.
\end{theorem}\

Before proving Theorem \ref{T1}, we note first that there is no loss of rigor by considering $C_{k}\left(\bar{\bt{P}}\right)$ in \eqref{E301} although $C_k(\cdot)$ depends only on $\bar{P}_{k}$. Then, since our focus is on asymptotically low power regime, both $\mathcal{R}\left(\bar{\bt{P}}\right)$ and $\mathcal{C}_{MAC}\left(\bar{\bt{P}}\right)$ collapse to the zero point. However, the characterization of the capacity region adopted in Theorem \ref{T1} is in the sense that the limit of the ratio between the achievable rate and the single user capacity is equal to 1 as $\bar{\bt{P}} \rightarrow0$ \cite{Caire2004a}.
\begin{IEEEproof}
We want to show that $\forall \;  \bm{\mu}=\left(\mu_1,\ldots,\mu_K\right) \, \in \mathbb{R}_{+}^{K}$, the point on boundary surface of the capacity region, $\bt{R}^{*}\left(\bm{\mu}\right)$ is independent of $\bm{\mu}$, i.e., all surfaces parameterizing the boundary surface of the capacity region intersect in exactly one point. This direct approach seems a bit complicated since the expression of $R^{*}_k$ in \cite[Theorem 3.16]{Tse1998} is somehow complicated. Instead, we adopt an information-theoretical approach to prove Theorem \ref{T1}. Clearly, the region on the right hand side (RHS) of \eqref{E301} is an upper bound on the MAC capacity region since the former is only constrained by single user capacities of all users. To show that all points in this set are achievable, it suffices to sow that the point $\bt{C}\left(\bar{\bt{P}}\right)=\left(C_{1}\left(\bar{P}_{1}\right),\ldots,C_{K}\left(\bar{P}_{K}\right)\right)$ is asymptotically achievable. For this purpose, let us compute $\bt{R}^{*}\left(\bm{\mu}\right)$ for $\mu_1=\mu_2=\ldots=\mu_K=1$, i.e., the point on the boundary of the capacity region that maximizes the sum rate. The $k$th components, $k=1,\ldots,K$, of this point is given by \cite{Tse1998}:
\begin{equation}\label{E303}
R^{*}_k=\int_{\lambda_{k}}^{\infty}\log\left(\frac{h}{\lambda_k}\right) \, \underset{i \neq k}{\prod}F_{i}\left(\frac{\lambda_{i}}{\lambda_{k}} \, h\right) \, f_k(h) \, \mathrm{d} h,
\end{equation}
where $f_k(\cdot)$ and $F_k(\cdot)$ denote the probability density function (pdf) of the channel gain power $\gamma_k$ and its cumulative distribution function (cdf), respectively; and where the constants $\lambda_k$'s satisfy
\begin{equation}\label{E305}
\bar{P}_k=\int_{\lambda_{k}}^{\infty}\left(\frac{1}{\lambda_k}-\frac{1}{h}\right) \, \underset{i \neq k}{\prod}F_{i}\left(\frac{\lambda_{i}}{\lambda_{k}} \, h\right) \, f_k(h) \, \mathrm{d} h.
\end{equation}
Let us define a function, say $G_k(x_1,\ldots,x_K)$, as the RHS of \eqref{E305}, i.e.,
\begin{equation}\label{E307}
G_k\left(x_1,\ldots,x_K\right)=\int_{x_k}^{\infty}\left(\frac{1}{x_k}-\frac{1}{h}\right) \, \underset{i \neq k}{\prod}F_{i}\left(\frac{x_{i}}{x_k} \, h\right) \, f_k(h) \, \mathrm{d} h,
\end{equation}
for positive $x_1,\ldots,x_K$. Note that $G_k\left(\lambda_1,\ldots,\lambda_K\right)=\bar{P}_k$, $k=1,\ldots,K$. Because each average power constraint depends on all $\lambda_i$'s, it is natural to consider that each $\lambda_{k}$ is a function of all $\bar{P}_i$'s, i.e., $\lambda_k=\lambda_k\left(\bar{P}_1,\ldots,\bar{P}_K\right)$.\footnote{Although rigorously speaking, in the context of the MAC, for any $k=1,\ldots,K$, $\lambda_{k}\left(\cdot\right)$ is a multivariate function of all $\bar{P}_{i}$, $i=1,\ldots,K$, i.e., $\lambda_k=\lambda_k\left(\bar{P}_1,\ldots,\bar{P}_K\right)$, we will find it convenient to denote it simply as $\lambda_{k}$ when there is no ambiguity.} We then claim the result in the following lemma.
\begin{lemma}\label{L1}
For the $\lambda_k$'s that satisfy the average power constraint \eqref{E305}, it holds that:
\begin{equation}\label{E309}
\underset{\bar{\bt{P}}\rightarrow \bt{0}}{\lim} \; \lambda_k\left(\bar{P}_1,\ldots,\bar{P}_K\right)=\infty
\end{equation}
for all $k=1,\ldots,K$. where $\bar{\bt{P}}\rightarrow \bt{0}$ stands for $\left(\bar{P}_1,\ldots,\bar{P}_K\right)\rightarrow \left(0,\ldots,0\right)$.
\end{lemma}
\begin{IEEEproof}
For convenience, the proof is presented in Appendix \ref{Appendix1}.
\end{IEEEproof}\
The intuition  behind \eqref{E309} is that at low power regime, $\lambda_{k}$ which has an interpretation of the power cost, converges toward infinity since at low power regime, the power becomes more expensive. This is true because the fading channels considered have infinite support.

Now, since $\forall h \in [\lambda_k,\infty)$, we have $F_{i}\left(\lambda_{i}\right) \leq F_{i}\left(\frac{\lambda_{i}}{\lambda_{k}} \, h\right) \leq 1$, the following inequalities hold true for all $\lambda_k >0$:
\begin{IEEEeqnarray}{rCl}
\underset{i \neq k}{\prod}F_{i}\left(\lambda_{i}\right) \; \int_{\lambda_{k}}^{\infty}\left(\frac{1}{\lambda_k}-\frac{1}{h}\right) \, f_k(h) \, \mathrm{d} h &\leq \nonumber \\
\int_{\lambda_{k}}^{\infty}\left(\frac{1}{\lambda_k}-\frac{1}{h}\right) \, \underset{i \neq k}{\prod}F_{i}\left(\frac{\lambda_{i}}{\lambda_{k}} \, h\right) \, f_k(h) \, \mathrm{d} h &\leq \nonumber \\
\int_{\lambda_{k}}^{\infty}\left(\frac{1}{\lambda_k}-\frac{1}{h}\right)  f_k(h) \, \mathrm{d} h, \label{E311}
\end{IEEEeqnarray}
or equivalently,
\begin{IEEEeqnarray}{rCrCl}
\underset{i \neq k}{\prod}F_{i}\left(\lambda_{i}\right)  &\leq&
\frac{\int_{\lambda_{k}}^{\infty}\left(\frac{1}{\lambda_k}-\frac{1}{h}\right) \, \underset{i \neq k}{\prod}F_{i}\left(\frac{\lambda_{i}}{\lambda_{k}} \, h\right) \, f_k(h) \, \mathrm{d} h}{\, \int_{\lambda_{k}}^{\infty}\left(\frac{1}{\lambda_k}-\frac{1}{h}\right) \, f_k(h) \, \mathrm{d} h} &\leq& 1. \label{E313}
\end{IEEEeqnarray}
Taking the limits as $\bar{\bt{P}} \rightarrow \bt{0}$ on both sides of \eqref{E313}, we establish that
\begin{equation}\label{E315}
\int_{\lambda_{k}}^{\infty}\left(\frac{1}{\lambda_k}-\frac{1}{h}\right) \, \underset{i \neq k}{\prod}F_{i}\left(\frac{\lambda_{i}}{\lambda_{k}} \, h\right) \, f_k(h) \, \mathrm{d} h \approx \, \int_{\lambda_{k}}^{\infty}\left(\frac{1}{\lambda_k}-\frac{1}{h}\right) \, f_k(h) \, \mathrm{d} h.
\end{equation}
Along similar lines, one can also prove that
\begin{equation}\label{E317}
\int_{\lambda_{k}}^{\infty}\log\left(\frac{h}{\lambda_k}\right) \, \underset{i \neq k}{\prod}F_{i}\left(\frac{\lambda_{i}}{\lambda_{k}} \, h\right) \, f_k(h) \, \mathrm{d} h \approx \int_{\lambda_{k}}^{\infty}\log\left(\frac{h}{\lambda_k}\right)  \, f_k(h) \, \mathrm{d} h.
\end{equation}
Note that the RHS of \eqref{E317} may be written as $\underset{\gamma_k}{\avr}\left[\log\left(1+\gamma_k \, \left[\frac{1}{\lambda_k}-\frac{1}{\gamma_k}\right]^{+}\right)\right]$ which takes clearly the form of an ergodic capacity expression due to the averaging over $\gamma_k$ and to the water-filling structure of the power inside the $\log(\cdot)$ function. Therefore, the RHS of \eqref{E317} describes the capacity of fading channel $k$ in function of $\lambda_k$ and not in function of $\bar{P}_k$ as we would have wished. But since $\lambda_{k}$ on the RHS of \eqref{E317} also satisfies \eqref{E315}, then we have
\begin{equation}\label{E319}
\bar{P}_k \approx \int_{\lambda_{k}}^{\infty}\left(\frac{1}{\lambda_k}-\frac{1}{h}\right) \, f_k(h) \, \mathrm{d} h
\end{equation}
due to \eqref{E305}. Hence, we conclude that at asymptotically low power regime, $R^{*}_k\left(\bar{P}_k\right) \approx  C_{k}\left(\bar{P}_{k}\right)$, for all $k=1,\ldots,K$. Finally, we highlight the fact that while $\lambda_k=\lambda_k\left(\bar{P}_1,\ldots,\bar{P}_K\right)$ for an arbitrary $\bar{P}_k$ values, \eqref{E319} stipulates that at asymptotically low power regime, the dependence on $\bar{P}_i$'s, $i \neq k$, breaks down so that $\lambda_k=\lambda_k\left(\bar{P}_k\right)$ only.
\end{IEEEproof}

It is worthwhile to mention that in order for our result to hold, the fading of different users need to be independent, not necessarily identically distributed. We also note that Theorem \ref{T1} emphasizes on how communication at low power regime can be energy efficient: at this regime, each user benefits from a single user performance as if others do not exist. The strategy to achieve this is the same as the one that maximizes the sum rate, a time-division multiple-access strategy where at most one user is allowed to transmit at any given state.

We note that since in the MAC scenario we perform water-filling over the maximum of users' channel gains $\underset{k=1,\ldots,K}{\max} \gamma_k$ to achieve the maximum sum-rate, whereas in the single-user case, we perform water-filling on $\gamma_{k}$, for each individual user, then these rates seem to be different a priori. However, the two strategies perform similarly at asymptotically low power regime as stated by Theorem \ref{T1}. At low power regime, since the power cost $\lambda_k \rightarrow \infty$, for all $k=1,\ldots,K$, if each user transmits only when its channel gain is ``extremely good" ($\gamma_{k} \geq \lambda_k$), then this channel gain is the strongest with high probability. That is, the probability that the channel gain $\gamma_k$ be bigger than all $\gamma_{j}$, $j\neq k$, given that $\gamma_k \geq \lambda_k$ converges to 1 as $\lambda_k \rightarrow \infty$, or equivalently as $\bar{P}_k \rightarrow 0$. The proof of Theorem \ref{T1} is somehow too technical and to get the feel of the insight behind Theorem \ref{T1}, we consider the following simple example.
\begin{example}\label{ex0}
Let us consider a symmetric MAC channel where all users have the same fading statistics and are constrained by the same average power constraint, i.e., $\bar{P}_k=\bar{P}$, for $k=1,\ldots,K$. As discussed above, TDMA achieves the maximum sum-rate. This yields a sum-rate equal to $R_{\sum}=\underset{\gamma_{max}}{\avr}\left[\log\left(1+\gamma_{max} \, P\left(\gamma_{max}\right)\right)\right]$, where $\gamma_{max}=\underset{k=1,\ldots,K}{\max} \gamma_k$ and the function $P(\cdot)$ is defined on $(0,\infty)$ by: $P(x)=\left[\frac{1}{\lambda}-\frac{1}{x}\right]^{+}$, with $\lambda$ chosen such that the average power constraint is satisfied with equality, i.e., $\underset{\gamma_{max}}{\avr}\left[P\left(\gamma_{max}\right)\right]=K \, \bar{P}$. Note that the factor $K$ in the later equation is due to the fact that each user has probability $1/K$ (symmetric fading) of being the one that has maximum channel gain. By symmetry, each user gets the rate $R_{k}=\frac{1}{K} \, \underset{\gamma_{max}}{\avr}\left[\log\left(1+\gamma_{max} \, P\left(\gamma_{max}\right)\right)\right]$ which can be lower-bounded as follows:
\begin{eqnarray}
R_{k}&=&\underset{\gamma_{k}}{\avr}\left[\log\left(1+\gamma_{k} \, P\left(\gamma_{k}\right)\right) \, \text{Prob}\left\{\gamma_j \leq \gamma_k, \; j\neq k\right\}\right] \label{E900} \\
     &\geq& \underset{\gamma_{k}}{\avr}\left[\log\left(1+\gamma_{k} \, P\left(\gamma_{k}\right)\right) \, \text{Prob}\left\{\gamma_j \leq \lambda, \; j\neq k\right\}\right] \label{E1000}\\
     &=&  \text{Prob}\left\{\gamma_j \leq \lambda, \; j\neq k\right\} \, \underset{\gamma_{k}}{\avr}\left[\log\left(1+\gamma_{k} \, P\left(\gamma_{k}\right)\right) \right]  \label{E1100}
\end{eqnarray}
where \eqref{E1000} follows because only $\gamma_k \geq \lambda$ matters in \eqref{E900}. Since $\lambda \rightarrow \infty$ as $\bar{P}\rightarrow 0$, then using \eqref{E1100}, we have:
\begin{equation}\label{E1200}
\underset{\bar{P}\rightarrow 0}{\lim} \; \frac{R_k}{\underset{\gamma_{k}}{\avr}\left[\log\left(1+\gamma_{k} \, P\left(\gamma_{k}\right)\right) \right]} \geq 1.
\end{equation}
From \eqref{E900}, we also have:
\begin{equation}\label{E1300}
\frac{R_k}{\underset{\gamma_{k}}{\avr}\left[\log\left(1+\gamma_{k} \, P\left(\gamma_{k}\right)\right) \right]} \leq 1.
\end{equation}
Combining \eqref{E1200} and \eqref{E1300}, we obtain:
\begin{equation}\label{E1400}
\underset{\bar{P}\rightarrow 0}{\lim} \;\frac{R_k}{\underset{\gamma_{k}}{\avr}\left[\log\left(1+\gamma_{k} \, P\left(\gamma_{k}\right)\right) \right]} = 1.
\end{equation}
Along similar steps, one can also show that:
\begin{equation}\label{E1500}
\underset{\bar{P}\rightarrow 0}{\lim} \;\frac{\bar{P}}{\underset{\gamma_{k}}{\avr}\left[P\left(\gamma_{k}\right) \right]} = 1.
\end{equation}
Note that \eqref{E1500} asserts that $\lambda$ satisfies the single user power constraint asymptotically at low power regime, i.e., $\underset{\gamma_{k}}{\avr}\left[P\left(\gamma_{k}\right) \right] \approx \bar{P}$; hence $C_k\left(\bar{P}\right) \approx \underset{\gamma_{k}}{\avr}\left[\log\left(1+\gamma_{k} \, P\left(\gamma_{k}\right)\right) \right]$, which combined with \eqref{E1400} yields $R_k \approx C_k\left(\bar{P}\right)$ as predicted by Theorem \ref{T1}.
\end{example}

\section{On-Off Power Control Achievable Rate}\label{S4}
It is well known that for the MAC channel, successive decoding is optimal \cite{Tse1998}. Here, we investigate the rate achieved by an on-off signaling. We show that at asymptotically low power regime, a simple on-off policy is optimal for a class of fading channels, in the sense that each user achieves the single user capacity bound.

In our scheme, each user communicates solely with the base station while other users are turned off so that the receiver can perform any optimal single-user decoding without the need of successive decoding. The power policy of each user is an appropriate on-off power control.

Before introducing the on-off signaling scheme, let us first introduce the class of fading channels we are interested in. Indeed, we focus on fading channels such that $\underset{t \rightarrow \infty}{\lim} \frac{t \, f_{k}\left(t\right)}{1-F_{k}\left(t\right)}$ exists (including infinity) and is strictly positive. For convenience, let $l_k$ be that limit, i.e., $l_k=\underset{t \rightarrow \infty}{\lim} \frac{t \, f_{k}\left(t\right)}{1-F_{k}\left(t\right)}$. Then, our focus is on fading channels such that $l_k>0$, $k=1,\ldots,K$. In reliability theory, the function $\zeta_k\left(t\right)=\frac{t \, f_{k}\left(t\right)}{1-F_{k}\left(t\right)}$ is known as Generalized Failure Rate (GFR) and the probability distributions with increasing GFR (also known as IGFR) satisfy $\underset{t \rightarrow \infty}{\lim} \; \zeta_k\left(t\right) >n$ if and only if $\avr{\left[\gamma_k^{n}\right]}$, for $n>0$, is finite \cite{Lariviere2006}. That is, for probability distributions with IGFR, having a finite mean is enough to show that $l_k>1$. We do not assume that the fading channels considered herein have an IGFR; however, we focus on fading channels with $l_k>0$ as the later condition is less restrictive than the IGFR property.

First, let us digress from the MAC setting and focus on a SISO point-to-point communication with perfect CSI-TR and claim the result in Theorem \ref{T2} that characterizes the capacity of this class of fading channels at low power regime.
\begin{theorem}\label{T2}
Let $l_k=\underset{t \rightarrow \infty}{\lim} \frac{t \, f_{k}\left(t\right)}{1-F_{k}\left(t\right)}$. For fading channels with $l_k>0$ (including $\infty$), the capacity of the discrete-time memoryless channel described by $y_k(n) = h_{k}(n) \; x_{k}(n) \, + \, v_k(n)$, $n=1,\ldots,\infty$, with perfect CSI-TR and under an average transmit power constraint $\bar{P}_k$, is given by
\begin{equation}\label{E4B1}
 C_k\left(\bar{P}_k\right) \approx \left(1+\frac{1}{l_k}\right) \, \lambda_k\left(\bar{P}_k\right) \; \bar{P}_k,
\end{equation}
where $\lambda_k\left(\bar{P}_k\right)$ is the water level corresponding to the capacity-achieving power profile. Furthermore, an on-off signaling defined by
\begin{equation}\label{E4B12}
P_{k}\left(\gamma_k\right)=
\begin{cases}
  \frac{\bar{P}_{k}}{1-F_{k}\left(\tau_k\right)}& \quad \text{if} \; \gamma_k \geq \tau_k \\
  0 & \quad \text{otherwise},
\end{cases}
\end{equation}
where $\tau_{k}=\left(1+\frac{1}{l_k}\right) \, \lambda_k\left(\bar{P}_k\right)$, is asymptotically capacity-achieving.
\end{theorem}
\begin{IEEEproof}
The proof is presented in Appendix \ref{Appendix3}
\end{IEEEproof}\

Theorem \ref{T2} provides an asymptotic expression of the capacity over a wide class of fading channels. Should a specific fading be chosen, the Lagrange multiplier $\lambda_k(\cdot)$ can be explicitly expressed in function of the average power $\bar{P}_k$. For instance, considering Rayleigh or the more general Nakagami-m fading, one can easily verify that $l_k \rightarrow \infty$, for both fading. Hence, they belong to the class of fading we are focusing in. Moreover, it has been shown in \cite{Rezki2012f} that $\lambda_k\left(\bar{P}_k\right) \approx \log{\left(\frac{1}{\bar{P}_k}\right)}$. Hence, using Theorem \ref{T2}, it can be seen immediately that the capacity scales essentially as $C_k(\bar{P}) \approx \bar{P}_k \, \log{\left(\frac{1}{\bar{P}_k}\right)}$ at low power regime. Note that this result is in full agreement with the ones established via different techniques in \cite{Borade2010a} and \cite{Rezki2012f}. Needless to recall that at asymptotically low power regime, the Lagrange multiplier $\lambda\left(\bar{P}_k\right)$ converges toward infinity. The proof is technical and can be found in Appendix \ref{Appendix3}. However, the intuition behind this fact can be easily understood by giving to $\lambda\left(\bar{P}_k\right)$ its economic interpretation as the power price. Since at low power regime, the power becomes more expensive, then it is natural that $\lambda\left(\bar{P}_k\right)$ increases indefinitely as $\bar{P}_k \rightarrow 0$. Note that for the class of fading channels in Theorem \ref{T2}, it has been shown in \cite{Verdu2002} that the ratio of the capacity and $\bar{P}_k$ goes to infinity. Theorem \ref{T2} describes this behavior more precisely stating that the capacity per unit power goes to infinity as $\left(1+\frac{1}{l_k}\right) \, \lambda_k\left(\bar{P}_k\right)$. This is in fact the capacity gain provided by perfect CSI-T, but the second statement of Theorem \ref{T2} clearly emphasizes that 1-bit feedback is actually enough for this to be true and proposes a general capacity-achieving on-off scheme that yields this gain. Finally, we note that in order for Theorem \ref{T2} to apply, we only require the $l_k$ to be positive. We do not claim that this condition is necessary, but we have verified that for all typical wireless channels encountered in the literature, this condition holds true. For instance, for Rayleigh, Rician and Nakagami fading channels, it can be easily verified that $l_k \rightarrow \infty$ and that the ergodic capacity is essentially equal to $K_0 \, \bar{P}_k \, \log{\left(\frac{1}{\bar{P}_k}\right)}$, where $K_0$ is a constant that depends on the fading statistics \cite{Borade2010a,Rezki2012b,Rezki2012f}.

We now return back to the MAC channel and propose a communication strategy that achieves the sum-rate capacity. For user $k$, let us consider an on-off power control scheme that transmits whenever $\gamma_k \geq \gamma_i$, $i \neq k$, and $\gamma_k \geq \tau_k$, where $\tau_{k}=\left(1+\frac{1}{l_k}\right) \, \lambda_k\left(\bar{P}_k\right)$ and $\lambda_k\left(\bar{P}_k\right)$ satisfies \eqref{E319}; with a fixed power equal to $Q_k$, and remains silent otherwise. To fulfill the average power constraint of each user, $Q_k$ is chosen such that:
\begin{equation}\label{E401}
Q_k=\frac{\bar{P}_k}{\int_{\tau_k}^{\infty} \, \underset{i \neq k}{\prod}F_{i}\left(\gamma\right) \, f_k(\gamma) \, \mathrm{d} \gamma}.
\end{equation}
That is, the instantaneous transmit power $P_k$ of user $k$ is defined by
\begin{equation}\label{E403A}
P_k\left(\gamma_1,\ldots,\gamma_K\right)=
\begin{cases}
  Q_k& \text{if} \; \gamma_k \geq \gamma_i \quad \text{and} \quad \gamma_k \geq \tau_k,\\
  0 & \text{otherwise}.
\end{cases}
\end{equation}
Before computing the achievable rate by this scheme, we note that because $\tau_k \rightarrow \infty$ as $\bar{P}_{k} \rightarrow 0$. We have:
\begin{IEEEeqnarray}{rCl}
Q_k &\leq& \frac{\bar{P}_k}{\underset{i \neq k}{\prod}F_{i}\left(\tau_k\right) \, \int_{\tau_k}^{\infty} \, f_k(\gamma) \, \mathrm{d} \gamma} \\
&\approx& \frac{\bar{P}_k}{1-F_k\left(\tau_k\right)}. \label{E403}
\end{IEEEeqnarray}
Since the RHS of \eqref{E403} converges to 0 as $\bar{P}_k \rightarrow 0$ as shown in Appendix \ref{Appendix3}, so does $Q_k$. We are now ready to compute the rate achieved by user $k$ as follows:
\begin{IEEEeqnarray}{rCl}
R_k&=& \idotsint{\left[\log{\left(1+ \gamma_k \; P_k\left(\gamma_1,\ldots,\gamma_K\right)\right)} \; \underset{i=1}{\prod^{K}} f_i(\gamma_i)\right] \; \mathrm{d} \gamma_1\ldots\mathrm{d} \gamma_K } \nonumber \\
   &=&  \int_{\tau_k}^{\infty}{\log{\left(1+ \gamma_k \; Q_k\right)} \;  f_k(\gamma_k) \; \underset{i \neq k}{\prod}F_i(\gamma_k) \; \mathrm{d} \gamma_k } \nonumber \\
             &\geq&  \int_{\tau_k}^{\infty}{\left(1-\epsilon\right) \; \gamma_k \; Q_k} \; f_k(\gamma_k) \;\underset{i \neq k}{\prod}F_i(\gamma_k) \; \mathrm{d} \gamma_k \label{E407} \\
             &\geq& \left(1-\epsilon\right) \; \tau_k \; \bar{P}_k, \label{E409}
\end{IEEEeqnarray}
where \eqref{E407} follows because $\underset{\bar{P}_k \rightarrow 0}{\lim} \, Q_k =0$, then $\forall \gamma_k \in [\tau_k,\infty)$, we have $\log{\left(1+\gamma_k \, Q_k\right)} \approx \gamma_k \, Q_k$ and thus $\left |\frac{\log{\left(1+\gamma_k \, Q_k\right)}}{\gamma_k \, Q_k}-1 \right \vert \leq \epsilon$, for all $\epsilon >0$, at sufficiently low $\bar{P}_k$. By taking the limits on both sides of \eqref{E409} as $\epsilon \rightarrow 0$, we establish that a rate equal to $\left(1+\frac{1}{l_k}\right) \, \lambda_k\left(\bar{P}_k\right)  \; \bar{P}_k$ is asymptotically achievable. This rate corresponds to the asymptotic capacity described by Theorem \ref{T2} and hence is the best rate we can achieve.

We finally note that in order to set the above achievability scheme, only $\log_2(K+1)$ feedback bits are required at each fading realization. These bits identify the user $k$, $k=1.\ldots,K$, if any, that is allowed to communicate at that fading realization. Furthermore, while at asymptotically low power regime, both TDMA and the proposed on-off scheme are optimal, as far as the optimality criterion is in the limit sense; we argue that the proposed on-off is easier to implement than TDMA. This is because in TDMA, the active user requires perfect CSI of its channel gain to perform a water-filling, whereas this is not necessary for the proposed scheme as a constant power is used instead.

\section{Capacity Region of the BC At Low Power Regime}\label{S5A}
In order to characterize the BC capacity region at low power regime, we utilize the established duality between the Gaussian MAC and BC so that we can deduce the BC capacity region from that of the MAC given in Theorem \ref{T1} \cite{Jindal2004}. Recall that the duality implies that the capacity region of the BC with power $\bar{P}$, is exactly equal to the capacity of the dual MAC, which has the same channel gains, and a sum power constraint of $\bar{P}$ across all $K$ transmitters. That is $\mathcal{C}_{BC}\left(\bar{P}\right)=\underset{\bm{1}\cdot \bm{P}=\bar{P}}{\bigcup}\mathcal{C}_{MAC}\left(\bm{P}\right)$, where $\bm{\alpha} \cdot \bm{P}=\underset{k=1}{\sum}^{K} \alpha_k \, P_k$ \cite{Jindal2004}. Our result is formalized in Theorem \ref{T3}.
\begin{theorem}\label{T3}
Let $\mathcal{R}^{'}\left(\bar{P}\right)$ be the region defined by:
\begin{equation}
\mathcal{R}^{'}\left(\bar{P}\right)=\left\{\bt{R}: \, R_{k} \, \leq \, C_{k}\left(\alpha_{k} \, \bar{P}\right), \; k=1,\ldots,K, \;   \underset{k=1}{\sum^{K}} \alpha_{k}= 1\right\},
\end{equation}
where $C_{k}\left(\alpha_{k} \, \bar{P}\right)$ is the single user capacity of user $k$, with average transmit power $\alpha_{k} \, \bar{P}$, and with perfect CSI-TR; and where $\alpha_k$'s are arbitrary positive coefficients. For the SISO BC described by (\ref{E201B}), the capacity region $\mathcal{C}_{BC}\left(\bar{P}\right)$ coincides with $\mathcal{R}^{'}\left(\bar{P}\right)$ at asymptotically low power regime. That is, all points in $\mathcal{R}^{'}\left(\bar{P}\right)$ are achievable and vice versa any point in $\mathcal{C}_{BC}\left(\bar{P}\right)$ is necessarily in $\mathcal{R}^{'}\left(\bar{P}\right)$ too.\footnote{Here again, as mentioned previously at the beginning of the proof of Theorem \ref{T1}, the achievability and the converse are in the sense that the limit of the ratio is equal to 1 as $\bar{P} \rightarrow 0$.}
\end{theorem}
\begin{IEEEproof}
To show that any point in $\mathcal{R}^{'}\left(\bar{P}\right)$ is achievable, we need only to focus on rates on the boundary of $\mathcal{R}^{'}\left(\bar{P}\right)$, i.e., the rates such that $R_{k} = C_{k}\left(\alpha_{k} \, \bar{P}\right)$, for all $k=1,\ldots,K$. Then, we know from Theorem \ref{T1} that the point with coordinates $\left(C_{1}\left(\alpha_{1} \, \bar{P}\right),\ldots,C_{K}\left(\alpha_{K} \, \bar{P}\right)\right)$ belongs to $\mathcal{C}_{MAC}\left(\bar{P} \, \bm{\alpha}\right)$. Since $\underset{k=1}{\sum^{K}}\alpha_{k} \, \bar{P}=\bar{P}$, then the point $\left(C_{1}\left(\alpha_{1} \, \bar{P}\right),\ldots,C_{K}\left(\alpha_{K} \, \bar{P}\right)\right)$ also belongs to $\mathcal{C}_{BC}\left(\bar{P} \right)$ by the duality between the Gaussian MAC and BC. This completes the achievability part.

To prove the converse, we only need to show that the points on the boundary of the BC capacity region necessarily belong to $\mathcal{R}^{'}\left(\bar{P}\right)$ too. To that end, let $\bm{R}$ be a point on the boundary of the BC capacity region. Then again, by the duality between the MAC and the BC, there exists a power policy $\bm{P}$ such $\bar{P}=\underset{k=1}{\sum^{K}}P_{k}$ and $\bm{R} \in \mathcal{C}_{MAC}\left(\bm{P}\right)$. For convenience, we let $P_{k} =\alpha_{k} \, \bar{P}$, for some positive $\alpha_{k}$'s, so that we have $\underset{k=1}{\sum^{K}} \alpha_{k}=1$. Furthermore, since $\bm{R}$ is on the boundary of the BC capacity region, then it is necessarily on the boundary of the $\mathcal{C}_{MAC}\left(\bm{P}\right)$ too, otherwise this would contradict the definition of the boundary curve. But, from Theorem \ref{T1} we know that the boundary of the MAC capacity region shrinks to a single point at low power regime, corresponding to the sum-rate maximizer and that $R_{k}\left(P_{k}\right) \approx C_{k}\left(P_{k}\right)$, or equivalently $R_{k}\left(\alpha_{k} \bar{P}\right) \approx C_{k}\left(\alpha_{k} \bar{P}\right)$. Therefore, the converse is also true and Theorem \ref{T3} is thus proved.
\end{IEEEproof}

Again, in order for Theorem \ref{T3} to hold, the fading of different users need not be identically distributed, only the independence is instrumental. We also note that from the proof above, since each point on the BC boundary region corresponds to a sum rate maximizer of a certain dual MAC, then to achieve a given point on the boundary of the BC capacity region, an optimal strategy is to use the MAC-BC transformation in each fading state to find a BC power policy that achieves the same average rate as the corresponding dual MAC \cite{Jindal2004}. That is, at any given state, the transmitter allocates all the power to at most one user as follows:
\begin{equation}\label{E5A02}
P\left(\bm{\gamma}\right)=
\begin{cases}
\left[\frac{1}{\lambda_{k}}-\frac{1}{\gamma_{k}}\right]^{+} & \text{if} \gamma_{k} > \frac{\lambda_k}{\lambda_j} \gamma_k, \; \forall j \\
0 & \text{else}
\end{cases}
\end{equation}
where $\lambda_k$'s are solution of $G_{k}\left(\bm{\lambda}\right)=\alpha_{k} \, \bar{P}$, for $k=1,\ldots,K$. Since at low power regime, $G_{k}\left(\bm{\lambda}\right) \approx \int_{\lambda_{k}}^{\infty}\left(\frac{1}{\lambda_k}-\frac{1}{h}\right) \, f_k(h) \, \mathrm{d} h$ due to \eqref{E315}, then $\lambda_k$'s can be obtained by solving a one-variable equation, i.e., $\int_{\lambda_{k}}^{\infty}\left(\frac{1}{\lambda_k}-\frac{1}{h}\right) \, f_k(h) \, \mathrm{d} h = \alpha_{k} \, \bar{P}$. Note that $\underset{\bm{\gamma}}{\avr}\left[P\left(\bm{\gamma}\right)\right]=\underset{k=1}{\sum^{K}}\alpha_{k} \, \bar{P}=\bar{P}$ confirming that the power policy given by \eqref{E5A02} satisfies the power constraint. It is then easy to check that rate achieved by each user using this strategy is asymptotically equal to $C_{k}\left(\alpha_k \, \bar{P}\right)$. Finally, we emphasize the fact that due to the concavity of $C_{k}\left(\cdot\right)$, we have $C_{k}\left(\alpha_k \, \bar{P}\right) \geq \alpha_{k} \,  C_{k}\left( \bar{P}\right)$, confirming also that the BC capacity region $\mathcal{R}^{'}\left(\bar{P}\right)$ contains the one corresponding to a time-sharing strategy where the available resource is exclusively allocated to user $k$ during a fraction of time $\alpha_{k}$. Consequently, time-sharing is asymptotically optimal if and only if $C_{k}\left(\alpha_k \, \bar{P}\right) \approx \alpha_{k} \,  C_{k}\left( \bar{P}\right)$, for all $k=1,\ldots,K$. Clearly, the later condition characterizes a class of fading where the power price $\lambda_{k}$ is asymptotically invariant to a scaling, i.e., $\lambda_{k}(\alpha \, \bar{P}) \approx \lambda_{k}\left(\bar{P}\right)$, for all $\alpha \in [0,1]$. To see this, let us first recall the expression of the single user capacity that depends only on the power constraint and not on the Lagrange multiplier $\lambda$ as established in Appendix \ref{Appendix3}:
\begin{equation}\label{E5A02AA}
C\left(\bar{P}\right)=\int_{0}^{\bar{P}}{G^{-1}(t) \, \mathrm{d}t},
\end{equation}
where $G^{-1}(\cdot)$ is the inverse function of $G(\cdot)$ and where the function $G(\cdot)$ is defined on $(0,\infty)$ by $G\left(\lambda\right)=\underset{\boldsymbol{\gamma}}{\avr}\left[\left[\frac{1}{\lambda}-\frac{1}{\gamma}\right]^{+}\right]$. We note that although not explicit, \eqref{E5A02AA} is interesting in the sense that it gives a simple expression of the capacity that does not depend on $\lambda$ and that is valid for an arbitrary $\bar{P}$ value.

Now, time-sharing is asymptotically optimal if and only if:
\begin{IEEEeqnarray}{rClll}
C_{k}\left(\alpha_k \, \bar{P}\right) \approx \alpha_{k} \,  C_{k}\left( \bar{P}\right) &\Leftrightarrow& \underset{\bar{P} \rightarrow 0}{\lim} \; \frac{C_{k}\left(\alpha_k \, \bar{P}\right) }{\alpha_{k} \,  C_{k}\left( \bar{P}\right)} &=&1 \\
                  &\Leftrightarrow& \underset{\bar{P} \rightarrow 0}{\lim} \; \frac{\int_{0}^{\alpha_k \, \bar{P}}{G_k^{-1}(t) \, \mathrm{d}t} }{\alpha_{k} \,  \int_{0}^{\bar{P}}{G_k^{-1}(t) \, \mathrm{d}t}}&=&1 \\
                  &\Leftrightarrow& \underset{\bar{P} \rightarrow 0}{\lim} \; \frac{\alpha_k \, G_k^{-1}(\alpha_{k} \, \bar{P})}{\alpha_{k} \, G_k^{-1}(\bar{P})} &=& 1 \label{E5A03}\\
                  &\Leftrightarrow& G_k^{-1}(\alpha_{k} \, \bar{P}) \approx G_k^{-1}( \bar{P}) &&\\
                  &\Leftrightarrow& \lambda_k(\alpha_{k} \, \bar{P}) \approx \lambda_k( \bar{P}),&& \label{E5A04}
\end{IEEEeqnarray}
where \eqref{E5A03} follows by l'H\^opital rule and \eqref{E5A04} is true by definition of the function $G_k(\cdot)$. Below we present an example of fading where this is actually the case.\\

\begin{example}[BC with independent Rayleigh fading channels]\label{ex1}
Consider a BC where the power channel gain for each user follows an exponential distribution $f_{k}\left(x\right)=\frac{1}{\bar{\gamma_{k}}} \, \text{exp}\left(-\frac{x}{\bar{\gamma_{k}}}\right)$. For these channels, it is shown that the single user capacity scales essentially as $C_{k}\left(\bar{P}\right) \approx \bar{\gamma_k} \, \bar{P}_k \log{\left(\frac{1}{\bar{P}_k}\right)}$ and that the Lagrange multiplier $\lambda_k\left(\bar{P}_k\right)$ scales essentially as $\lambda_k\left(\bar{P}_k\right) \approx  \log{\left(\frac{1}{\bar{P}_k}\right)} -2 \, \log\left(\log{\left(\frac{1}{\bar{P}_k}\right)}\right)$ \cite{Borade2010a}. We note that for these channels, $\lambda_k\left(\alpha \, \bar{P}_k\right) \approx  \log{\left(\frac{1}{\alpha \, \bar{P}_k}\right)} -2 \, \log\left(\log{\left(\frac{1}{\alpha \, \bar{P}_k}\right)}\right) \approx  \log{\left(\frac{1}{\bar{P}_k}\right)} -2 \, \log\left(\log{\left(\frac{1}{ \bar{P}_k}\right)}\right) \approx \lambda_k\left( \bar{P}_k\right)$. Hence, for these popular fading channels, time-sharing among users achieves the boundary of the capacity region $\mathcal{R}^{'}\left(\bar{P}\right)$.
\end{example}\

However, property \eqref{E5A04} does not hold in general as evidenced by the following example.\\

\begin{example}[BC with i.i.d. log-logistic fading channels]\label{ex2}
Consider a BC where the power channel gain for each user follows a log-logistic distribution $f_{k}\left(x\right)=\frac{1}{\left(1+x\right)^2}$. For these channels, using Theorem \ref{T2}, it can be easily verified that the single user capacity scales essentially as $C_{k}\left(\bar{P}_k\right) \approx 2 \, \lambda_{k}\left(\bar{P}_k\right) \, \bar{P}_k$. On the other hand, it can be shown that the Lagrange multiplier $\lambda_{k}\left(\bar{P}_k\right)\approx \frac{1}{\sqrt{2 \bar{P}_k}}$. Therefore, property \eqref{E5A04} does not hold in this case since $\lambda_{k}\left(\alpha \, \bar{P}_k\right) \approx  \frac{1}{\sqrt{2 \alpha \, \bar{P}_k}} \approx \frac{1}{\sqrt{\alpha}} \, \lambda_{k}\left(\bar{P}_k\right)$. Hence, for these fading channels, time-sharing among users is strictly suboptimal.
\end{example}

\section{Numerical Results and Discussion}\label{S5}
\begin{figure}
  \centering
    \includegraphics[scale=0.27]{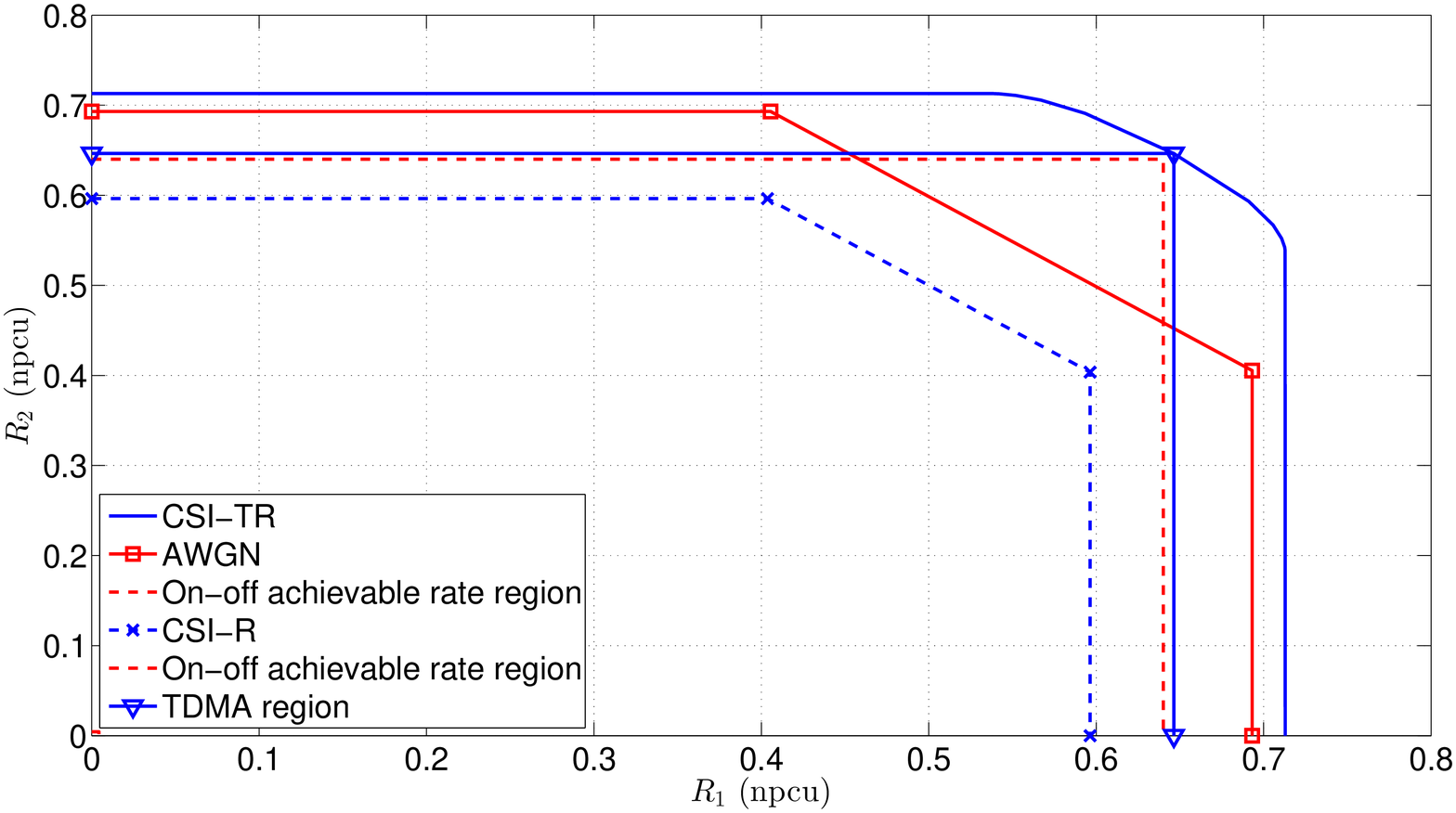}
        \caption{Capacity region and on-off achievable rate region of a 2-user MAC Rayleigh fading channel, with $\bar{P}_1=\bar{P}2=0$ dB. }
    \label{F2}
\end{figure}

\begin{figure}
  \centering
    \includegraphics[scale=0.27]{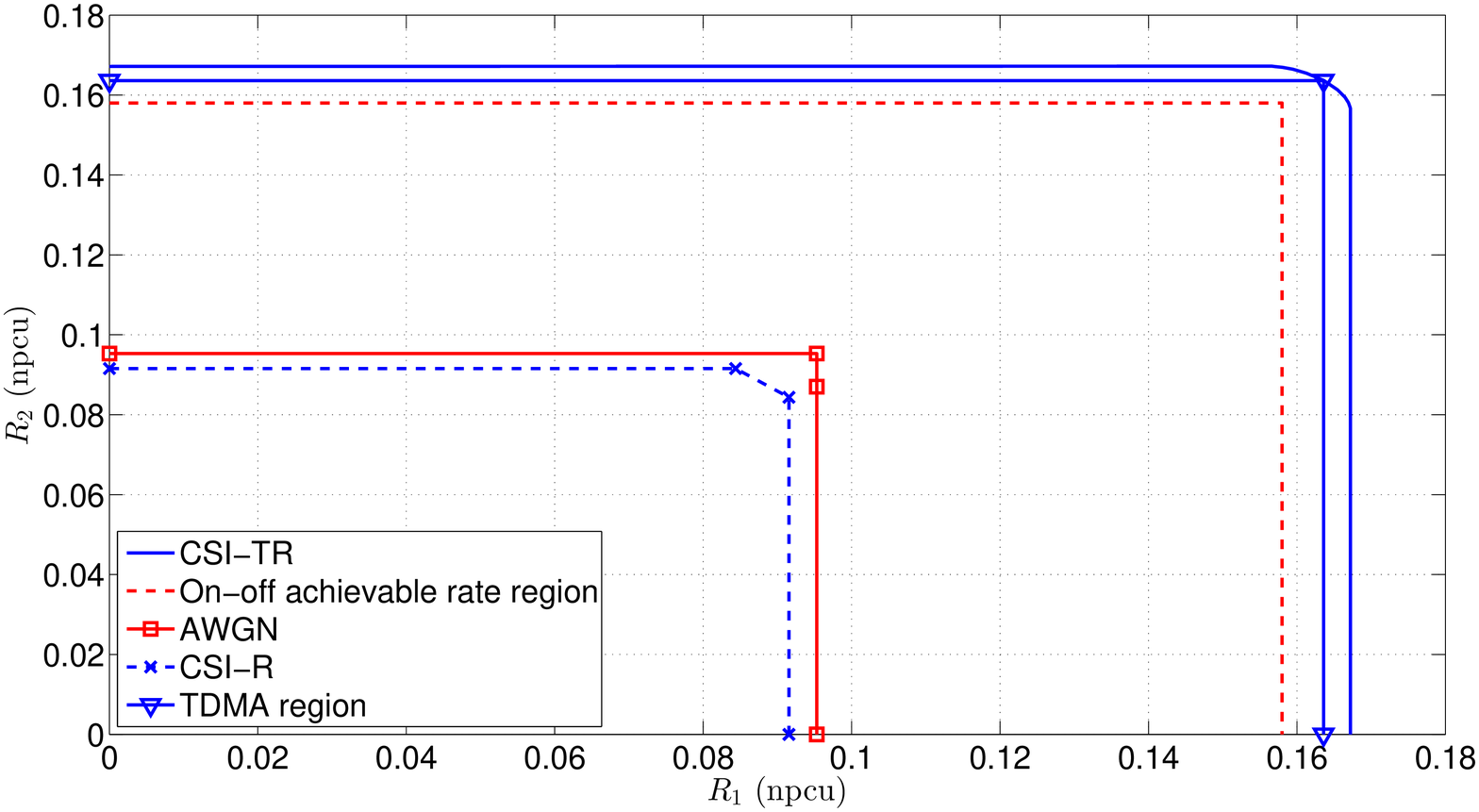}
        \caption{Capacity region and on-off achievable rate region of a 2-user MAC Rayleigh fading channel, with $\bar{P}_1=\bar{P}2=-10$ dB. }
    \label{F3}
\end{figure}

\begin{figure}
  \centering
    \includegraphics[scale=0.27]{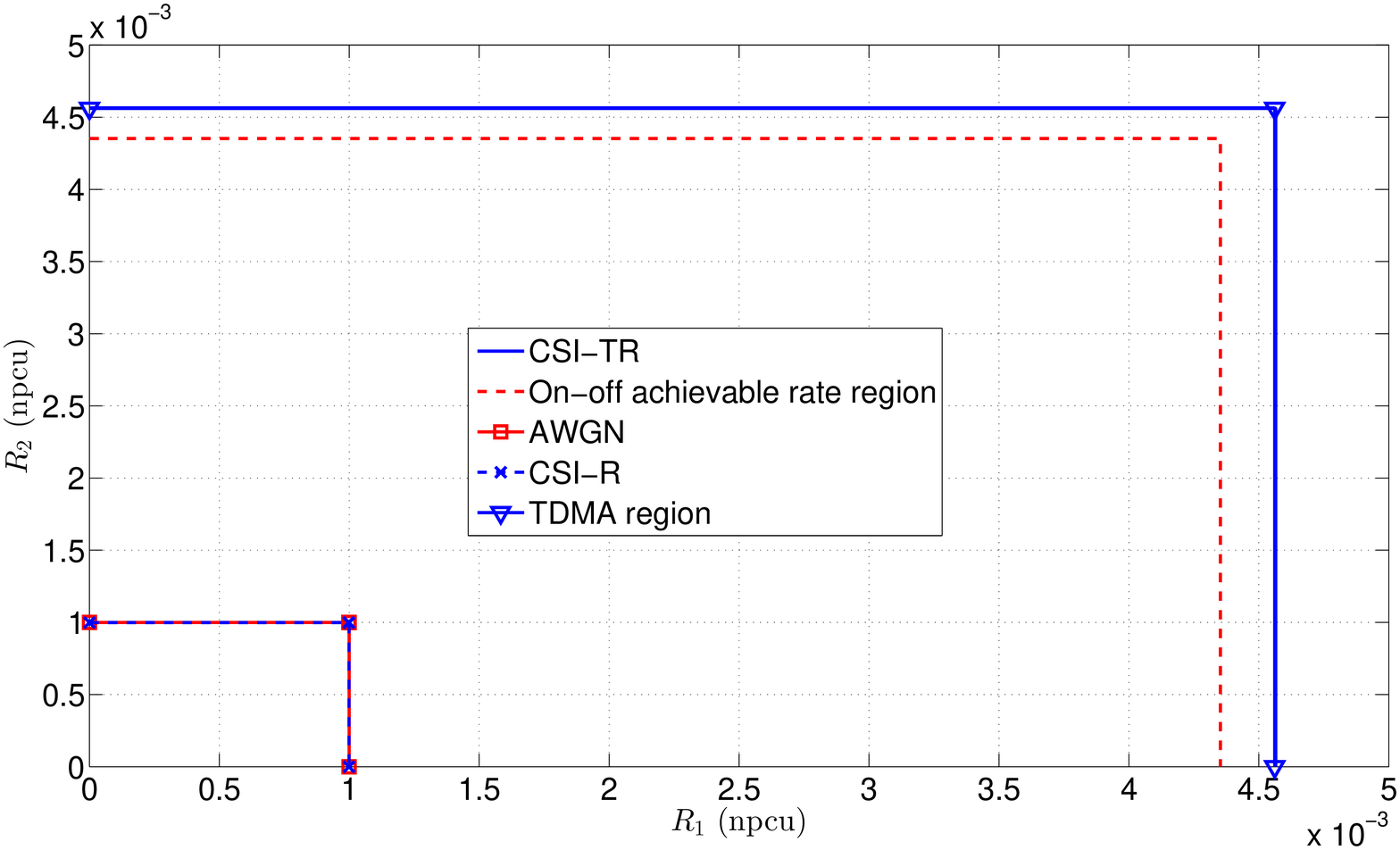}
        \caption{Capacity region and on-off achievable rate region of a 2-user MAC Rayleigh fading channel, with $\bar{P}_1=\bar{P}_2=-30$ dB. }
    \label{F4}
\end{figure}

\begin{figure}
  \centering
    \includegraphics[scale=0.27]{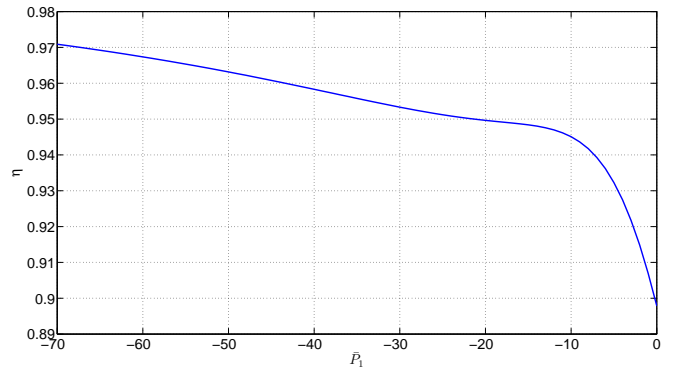}
        \caption{Ratio $\eta$, between the rate achieved by the proposed on-off scheme and the single user capacity versus $\bar{P}_1$, for a 2-user MAC Rayleigh fading channel.}
    \label{F5bis}
\end{figure}

\begin{figure}
  \centering
    \includegraphics[scale=0.27]{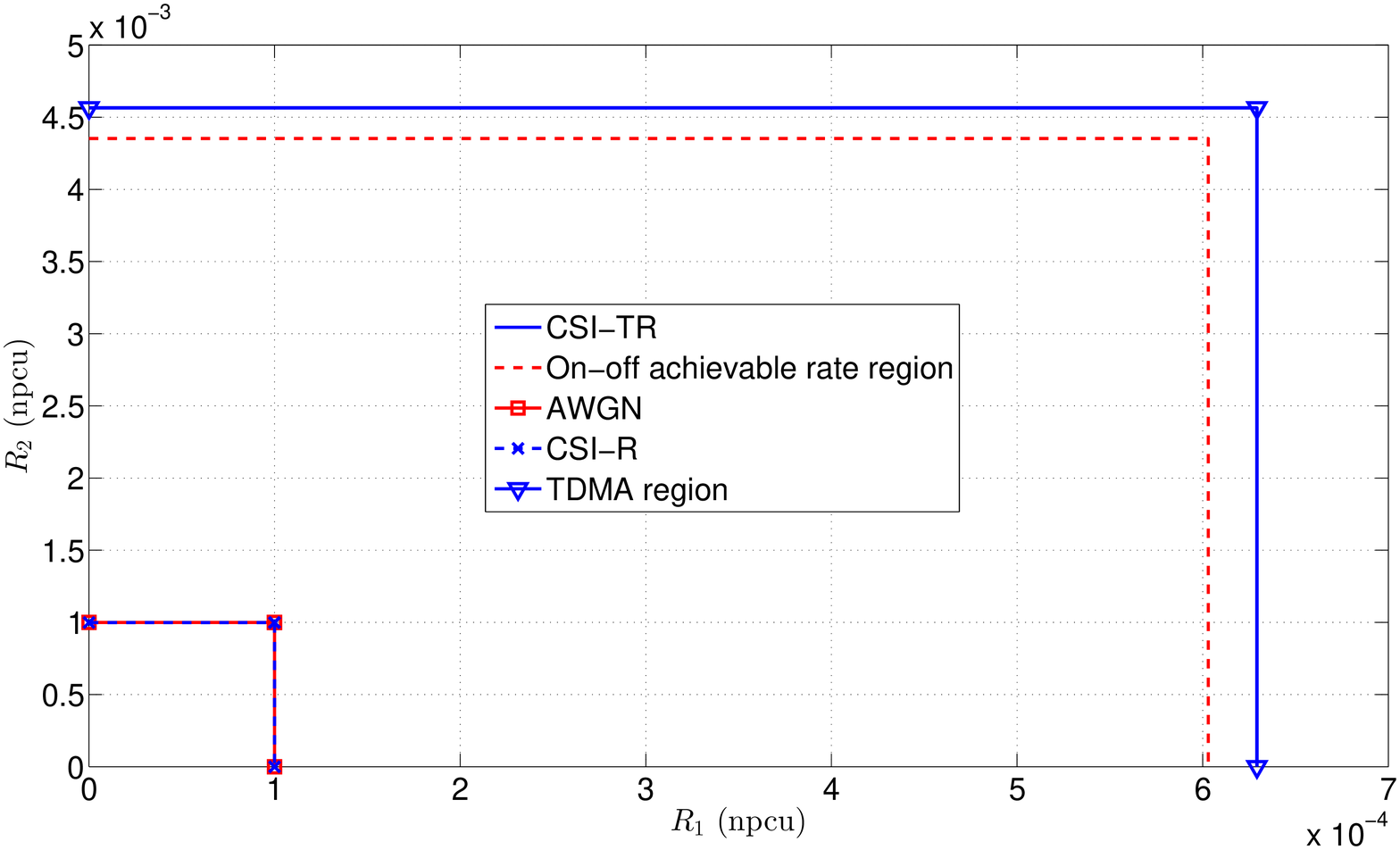}
        \caption{Capacity region and on-off achievable rate region of a 2-user MAC Rayleigh fading channel, with $\bar{P}_1=-40$ dB and $\bar{P_2}=-30$ dB. }
    \label{F5}
\end{figure}

\begin{figure}
  \centering
    \includegraphics[scale=0.27]{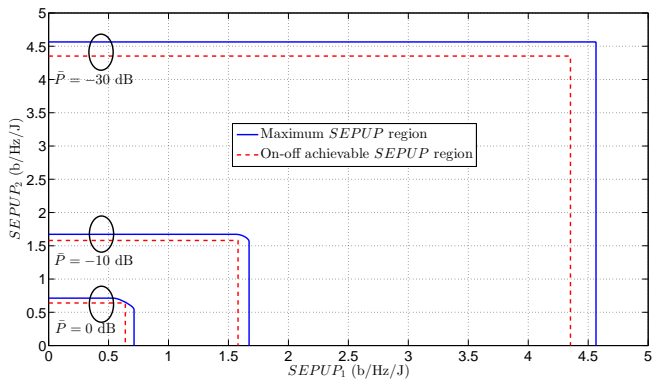}
        \caption{Spectral efficiency per unit power (SEPUP) region for a 2-user MAC Rayleigh fading channel for different $\bar{P}$ values.}
    \label{F6bis}
\end{figure}

We present numerical results for a 2-user MAC channel where both users undergo independent Rayleigh fading channels. In all figures, we depict the actual throughput capacity region obtained numerically using the characterization in \cite[Theorem 3.16]{Tse1998}, TDMA's achievable rate-region, the throughput achievable region by the proposed on-off scheme, the capacity region of an additive white Gaussian noise (AWGN) MAC channel and that of a Rayleigh fading MAC channel with CSI-R only. We note first that CSI-T provides a tremendous capacity gain over CSI-R only, especially at low power regime as can be seen in Fig. \ref{F2}, Fig. \ref{F3}, Fig. \ref{F4} and Fig. \ref{F5}. Furthermore, progressively from Fig. \ref{F2} to Fig. \ref{F4}, the throughput capacity region converges to a square, since the transmit powers of the 2 users are equal. In the previous figures, both TDMA's rate-region and the rate-region of the proposed on-off scheme converge toward the capacity region at low power regime. However, the convergence of TDMA's region toward the capacity region is faster than that of the proposed scheme. Note that the ratio between the maximum on-off achievable rate and the single user capacity bound of the MAC capacity region increases from 0.9 for $\bar{P}_1=\bar{P}_2=0$ dB in Fig. \ref{F2}, to 0.95 for $\bar{P}_1=\bar{P}_2=-30$ dB in Fig. \ref{F4}. Decreasing $\bar{P}_1$ and $\bar{P}_2$ further, results in the convergence of the ratio to 1. For a better illustration as to how the proposed on-off scheme approaches the MAC capacity region at low-power regime, we provide in Fig. \ref{F5bis} a plot that describes the ratio $\eta$, between the rate achieved by the proposed on-off scheme and the single user capacity, versus $\bar{P}_1$. Clearly, since the rate region achieved by the on-off scheme is certainly a multi-dimensional rectangle, then $\eta$ describes how close is the on-off achievable rate region to the MAC capacity region. As can be seen in Fig. \ref{F5bis}, below -10 dB, the on-off scheme has already achieved more than $94\%$ of the single user capacity bound. Similar observations can be noted from Fig. \ref{F5} for $\bar{P}_1=-30$ dB and $\bar{P}_2=-40$ dB. It is also worthwhile to observe that the ratio between the achievable rate and the average power increases with the power budget. This implies that a better spectral efficiency is achieved at low power regime. Since this ratio has dimension of spectral efficiency per unit power (bits per second per Hertz over Joules per second), we found it more convenient to call it spectral efficiency per unit power (SEPUP), the unit of which is bits per Hertz per Joule (b/Hz/J). Furthermore, one may define a SEPUP region for the MAC that describes the maximum spectral efficiency per unit power users can achieve. The SEPUP region follows in a straightforward manner from the capacity region of the MAC channel. Indeed, in a $K$-user MAC channel, given a transmit power budget $\left(\bar{P}_{1},\ldots,\bar{P}_{K}\right)$ of the users, each point of the capacity region defined by $\left(R_{1},\ldots,R_{K}\right)$, specifies a point of the SEPUP region, say $\left(SEPUP_{1},\ldots,SEPUP_{K}\right)$, such that $SEPUP_{k}=\frac{R_{k}}{\bar{P}_{k}}$. Analogously to single-user communications, the low-power regime provides the best (highest) SEPUP, or equivalently the best (lowest) minimum energy per bit as defined in \cite{Verdu2002}. Figure \ref{F6bis} depicts the SEPUP region for a 2-user MAC Rayleigh fading channel versus for different $\bar{P}$ values, where $\bar{P}_1=\bar{P}_2=\bar{P}$. Also shown in Fig. \ref{F6bis} is the SEPUP region achieved by the proposed on-off scheme. As can be seen in Fig. \ref{F6bis}, the SEPUP region gets larger as $\bar{P}$ decreases. Albeit, we have presented results where both users' channels undergo Rayleigh fading channel, we emphasize that the same trend holds as long as both channels are independent, but not necessarily identically distributed.\

\begin{figure}
  \centering
    \includegraphics[scale=0.27]{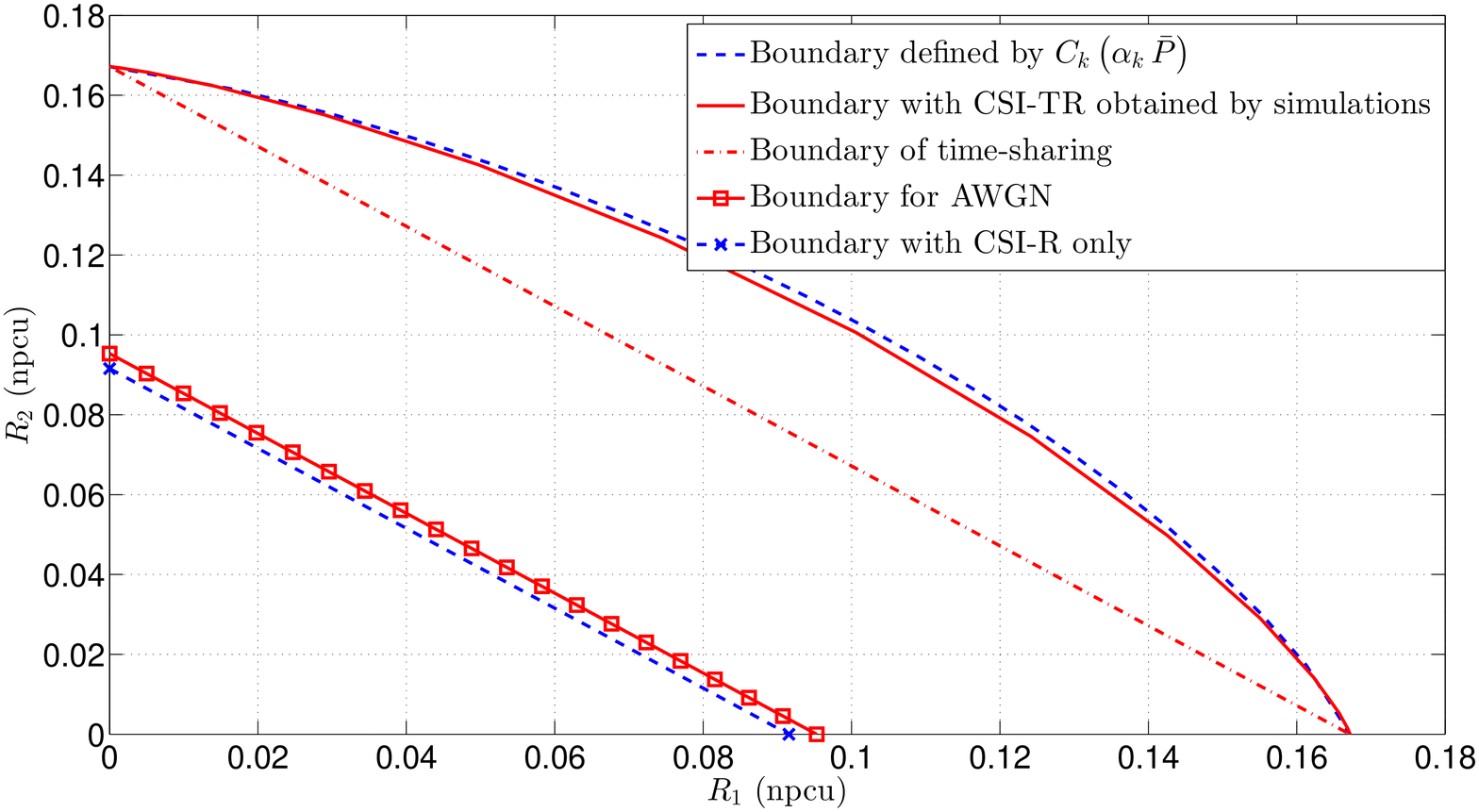}
        \caption{Capacity region of a 2-user BC Rayleigh fading channel, with $\bar{P}=-10$ dB. }
    \label{F6}
\end{figure}

\begin{figure}
  \centering
    \includegraphics[scale=0.27]{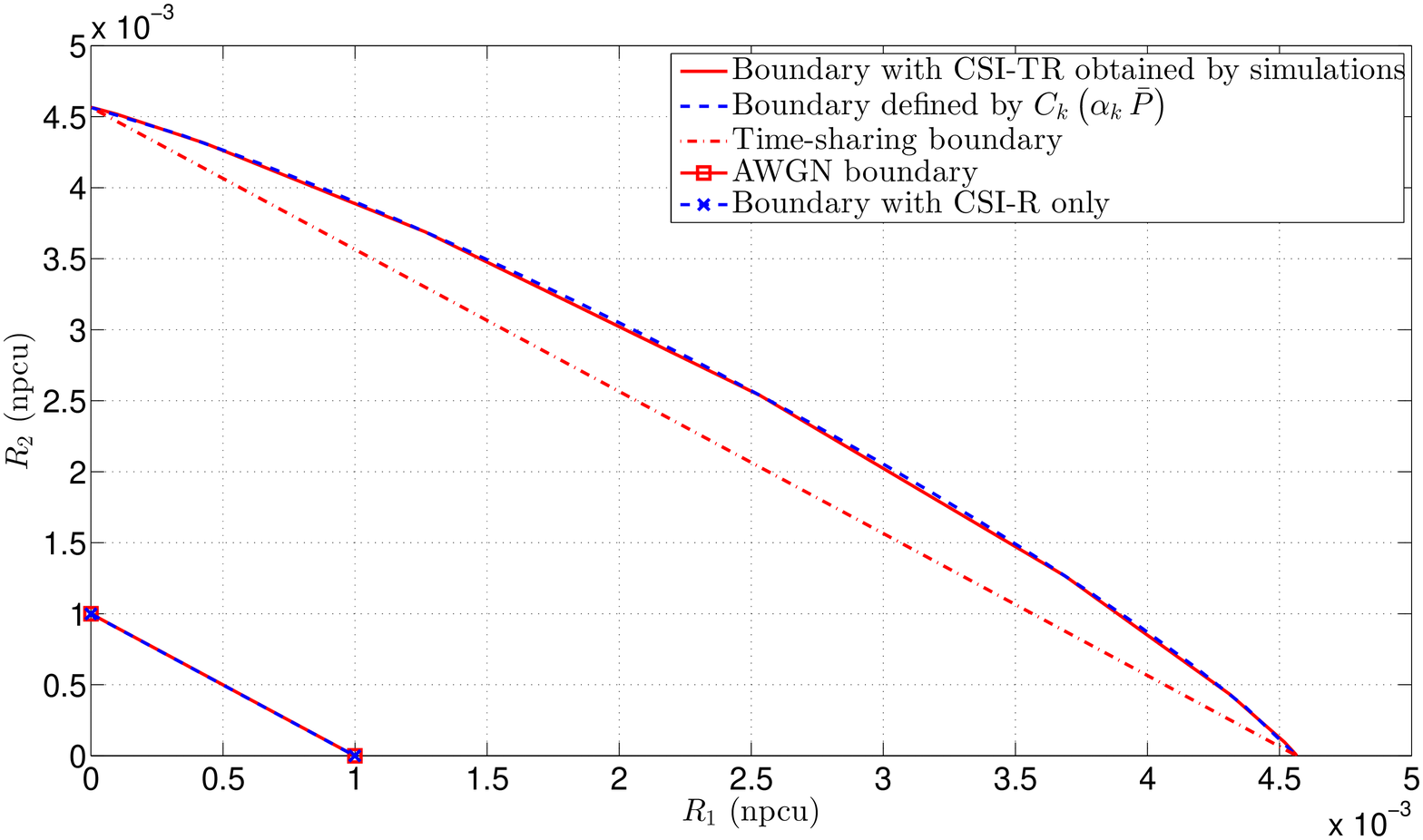}
        \caption{Capacity region of a 2-user BC Rayleigh fading channel, with $\bar{P}=-30$ dB. }
    \label{F7}
\end{figure}

\begin{figure}
  \centering
    \includegraphics[scale=0.27]{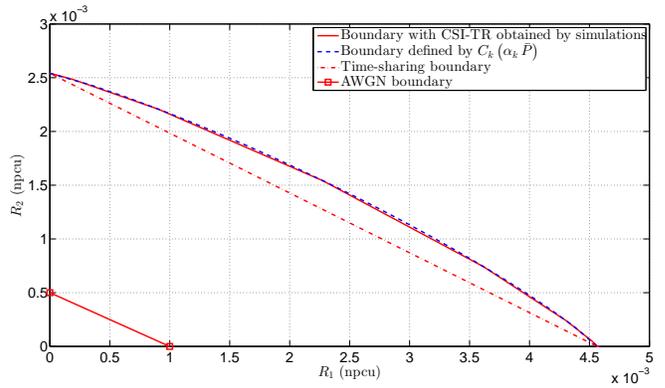}
        \caption{Capacity region of a 2-user BC Rayleigh fading channel, with $\bar{P}=-30$ dB and the difference of the power gains of the two users equal to $3$ dB. }
    \label{F8}
\end{figure}

\begin{figure}
  \centering
    \includegraphics[scale=0.27]{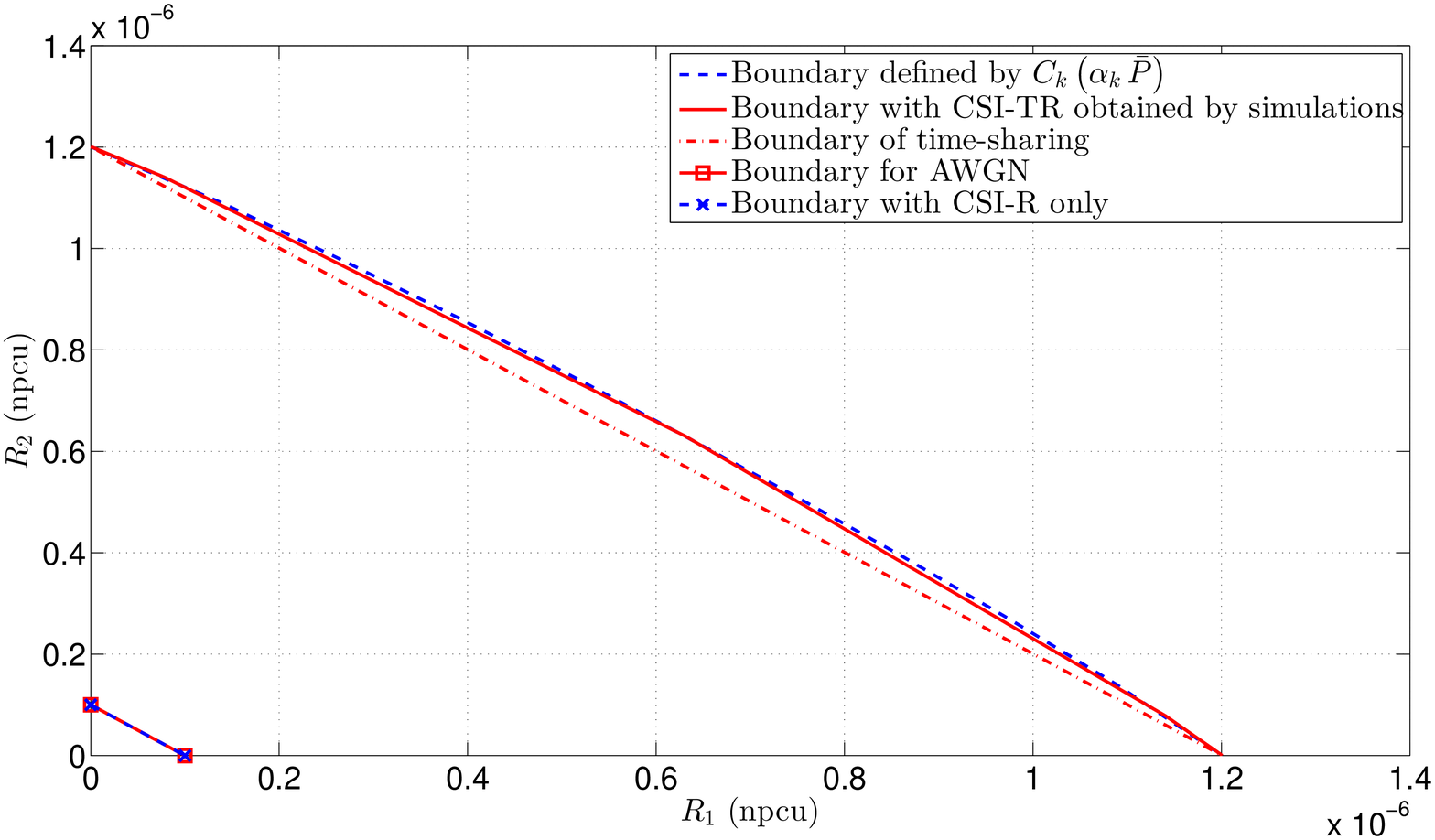}
        \caption{Capacity region of a 2-user BC Rayleigh fading channel, with $\bar{P}=-70$ dB. }
    \label{F9}
\end{figure}

\begin{figure}
  \centering
    \includegraphics[scale=0.27]{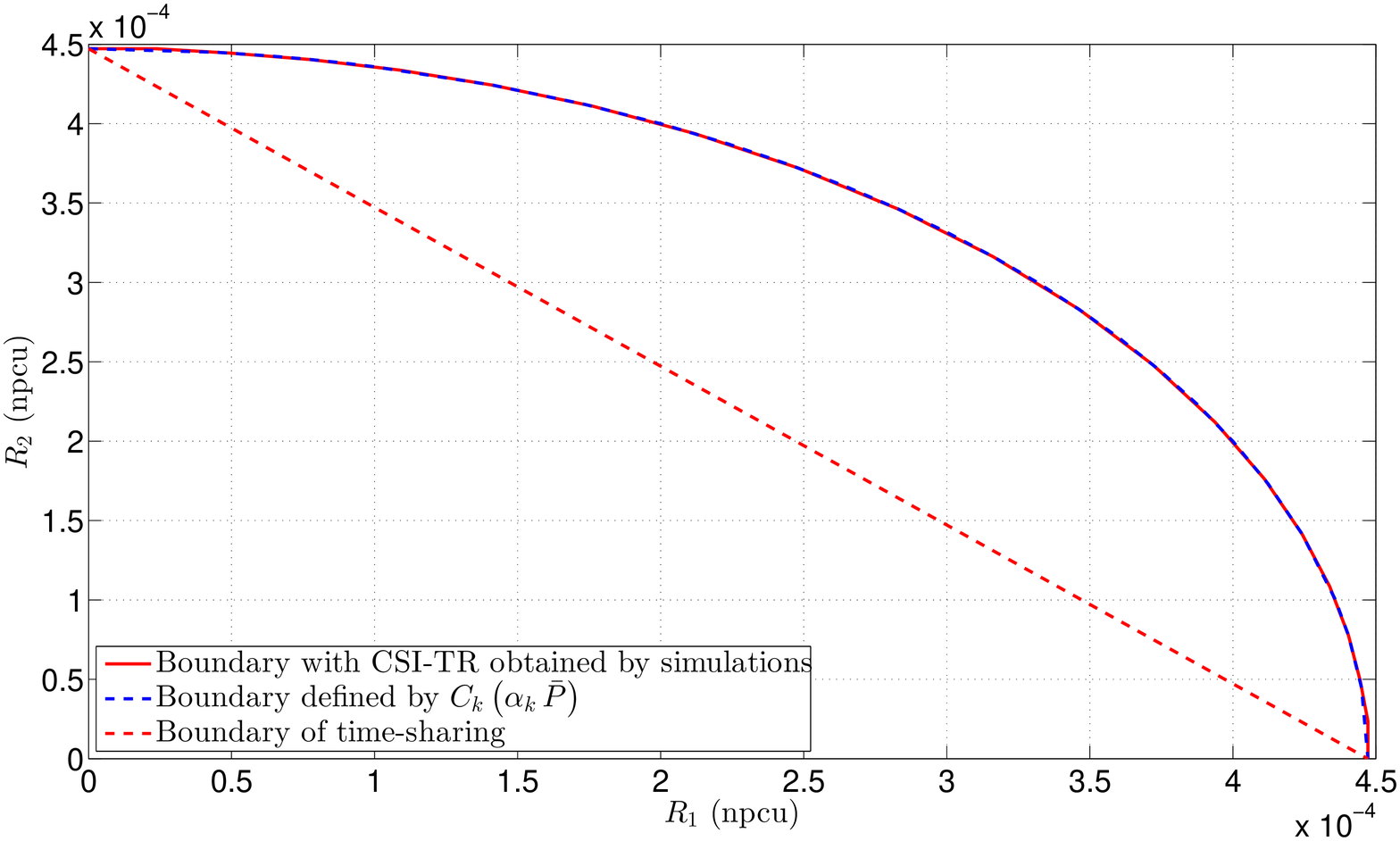}
        \caption{Capacity region of a 2-user BC log-logistic fading channel, with $\bar{P}=-70$ dB. }
    \label{F10}
\end{figure}

For the BC, Fig. \ref{F6}, Fig. \ref{F7}, Fig. \ref{F8} and Fig. \ref{F9} display the capacity region obtained numerically, the one given by Theorem \ref{T3} along with time-sharing achievable region, for $\bar{P}=-10$ dB, $\bar{P}=-30$ dB, $\bar{P}=-30$ dB with unequal power channel gain where the difference of the power gains of the two users is equal to $3$ dB, and $\bar{P}=-70$ dB, respectively. Recall that while the capacity region of the fading BC with CSI-R only is generally not know, it is actually known in the symmetric fading case considered in Fig. \ref{F6}, Fig. \ref{F7}, and Fig. \ref{F9}. Here again, beyond the huge capacity gain provided by CSI-T over CSI-R only, we note in all figures the accuracy of the characterization in Theorem \ref{T3}. Although time-sharing achievable region is strictly suboptimal at $-10$ dB, it converges slowly to the capacity region as shown in Fig. \ref{F9}. However, this is not true in general as established by \eqref{E5A04}. Indeed, Fig. \ref{F10} depicts the BC capacity region, the characterization in Theorem \ref{T3} and time-sharing performance for log-logistic fading as described by Example \ref{ex2}. Expectedly, the characterization in Theorem \ref{T3} is still very accurate as it matches perfectly well the actual BC capacity region, but time-sharing is strictly sub-optimal in this case even at $\bar{P}=-70$ dB.

\section{Conclusion}\label{S6}
We have analyzed the throughput capacity region of the MAC fading channel with perfect CSI at the transmitters and at the receiver at low power regime. While the capacity region has a polymatroid structure at arbitrary power regime, We have shown that it is simply a multidimensional rectangle at asymptotically low power regime and that each user can achieve a single user performance as if others do not exist. We have also proposed a simple on-off scheme, computed its achievable rate and characterized a class of fading channels for which the proposed scheme is asymptotically optimal. We have deduced the BC capacity region using the duality and provided a simple asymptotic characterization of the BC capacity region as well. A class of fading channels for which time-sharing is asymptotically optimal has been identified.



\appendices
\section{Proof of Lemma \ref{L1}}\label{Appendix1}
We recall that $\lambda_k$'s, $k=1,\ldots,K$, are solution of \eqref{E305} and thus we have $G_k\left(\lambda_1,\ldots,\lambda_K\right)=\bar{P}_k$. We also note that $\forall k$, $\lambda_{k} >0$, otherwise if there exists a certain $k_0$ such that $\lambda_{k_0}=0$, then $\bar{P}_{k_{0}} \rightarrow \infty$ according to \eqref{E305} which contradicts the fact that user's average powers are finite. Let $\lambda_{m}=\underset{k=1,\ldots,K}{\min} \, \lambda_k$, then the following inequalities hold true:
\begin{IEEEeqnarray}{rCl}
\bar{P}_m &=& \int_{\lambda_{m}}^{\infty}\left(\frac{1}{\lambda_m}-\frac{1}{\gamma}\right) \, \underset{i \neq m}{\prod}F_{i}\left(\frac{\lambda_{i}}{\lambda_{m}} \, \gamma\right) \, f_m(\gamma) \, \mathrm{d} \gamma \nonumber \\
    &\geq& \int_{\lambda_{m}}^{\infty}\left(\frac{1}{\lambda_m}-\frac{1}{\gamma}\right) \, \underset{i \neq m}{\prod}F_{i}\left(\lambda_{i}\right) \, f_m(\gamma) \, \mathrm{d} \gamma \nonumber \\
    &\geq& \underset{i \neq m}{\prod}F_{i}\left(\lambda_{m}\right) \;  \int_{\lambda_{m}}^{\infty}\left(\frac{1}{\lambda_m}-\frac{1}{\gamma}\right)  f_m(\gamma) \, \mathrm{d} \gamma \label{A1E1}.
\end{IEEEeqnarray}
The RHS of \eqref{A1E1} is positive and $\underset{\bar{\bm{P}}\rightarrow \bm{0}}{\lim} \bar{P}_m=0$, then by the Sandwich Theorem, we have
\begin{equation}\label{A1E5}
\underset{\bar{\bm{P}}\rightarrow \bm{0}}{\lim} \; \left(\underset{i \neq m}{\prod}F_{i}\left(\lambda_{m}\right) \;  \int_{\lambda_{m}}^{\infty}\left(\frac{1}{\lambda_m}-\frac{1}{\gamma}\right)  f_m(\gamma) \, \mathrm{d} \gamma\right)=0.
 \end{equation}
Since $\lambda_m >0$, then \eqref{A1E5} implies necessarily  that $\underset{\bar{\bm{P}}\rightarrow \bm{0}}{\lim} \int_{\lambda_{m}}^{\infty}\left(\frac{1}{\lambda_m}-\frac{1}{\gamma}\right) f_m(\gamma) \, \mathrm{d} \gamma=0$. The target in the later equation is nothing but the $G_m(\cdot)$ function defined in \cite{Rezki2012f} which is continuous and monotonically decreasing with $\underset{x\rightarrow\infty}{\lim} \, G_m(x)=0$. Hence, $\underset{\bar{\bm{P}}\rightarrow \bm{0}}{\lim} \, \lambda_{m}=\infty$. In summary, we have shown that the $\lambda_{m}=\underset{k=1,\ldots,K}{\min} \, \lambda_k \, \rightarrow \infty$ as $\bar{\bm{P}} \rightarrow \bm{0}$, we conclude that $\forall k=1,\ldots,K$, we have $\underset{\bar{\bt{P}}\rightarrow \bt{0}}{\lim} \; \lambda_k\left(\bar{P}_1,\ldots,\bar{P}_K\right)=\infty$ and Lemma \ref{L1} is thus proved.

\section{Proof of Theorem \ref{T2}}\label{Appendix3}
We first prove the asymptotic capacity expression given by \eqref{E4B1}. For the channel described by $y_k(n) = h_{k}(n) \; x_{k}(n) \, + \, v_k(n)$, $n=1,\ldots,\infty$, with perfect CSI-TR and under an average transmit power constraint $\bar{P}_k$, it is well-known that the instantaneous optimal power, $P_k\left(\gamma\right)$, is a water-filling policy given by \cite{Goldsmith1997a}: $P_k\left(\gamma_k\right)=\left[\frac{1}{\lambda_k\left(\bar{P}_k\right)}-\frac{1}{\gamma_k}\right]^{+}$ where $\lambda_k\left(\bar{P}_k\right)$ is the Lagrange multiplier obtained by satisfying the average power constraint with equality. That is, $G_k\left(\lambda_k\left(\bar{P}_k\right)\right)\stackrel{\Delta}{=}\int_{\lambda_k\left(\bar{P}_k\right)}^{\infty}\left(\frac{1}{\lambda_k\left(\bar{P}_k\right)}-\frac{1}{\gamma_k}\right) \, f_k\left(\gamma_k\right) \mathrm{d} \gamma_k=\bar{P}_k$. The capacity is then obtained by averaging $\log{\left(1+P_k(\gamma_k) \; \gamma_k\right)}$ and is given by
\begin{equation}\label{E4B2}
C_k\left(\lambda_k\left(\bar{P}_k\right)\right)=
\int_{\lambda_k\left(\bar{P}_k\right)}^{\infty}
{\log{\left(\frac{t}{\lambda_k\left(\bar{P}_k\right)}\right)f_{k}(t) \, \mathrm{d}t}}.
\end{equation}
Now, let us summarize few properties of the function $G_{k}(\cdot)$ in Lemma \ref{L3}.
\begin{lemma}\label{L3}
The function $G_k\left(\cdot\right)$ defined by $G_k\left(t\right)=\int_{t}^{\infty}\left(\frac{1}{t}-\frac{1}{\gamma_k}\right) \, f_k\left(\gamma_k\right) \mathrm{d} \gamma_k$ is i) continuous and positive definite; ii) strictly monotonically decreasing; iii) invertible on its domain, iv) $\underset{t \rightarrow \infty}{\lim} \; G_k\left(t\right)=0$, v) $\underset{t \rightarrow \infty} {\lim} \; t \, G_k\left(t\right)=0$ and vi) if $l_k=\underset{t \rightarrow \infty}{\lim} \; \frac{t \, f_{k}\left(t\right)}{1-F_{k}\left(t\right)}$, then $\underset{t \rightarrow \infty} {\lim} \; \frac{t \; G_k\left(t\right)}{1-F_k(t)}=\frac{1}{1+l_k}$.
\end{lemma}
\begin{IEEEproof}
The proof of i), ii) and iii) is presented in \cite{Rezki2012f}. The remaining parts of the proof are as follows.
\begin{itemize}
\item Proof of iv)
\end{itemize}
Since $\forall t \in (0,\infty)$, we have $0 < G_k\left(t\right) < \frac{1-F_k\left(t\right)}{t} < \frac{1}{t}$, then by the Squeeze Theorem, $\underset{t\rightarrow \infty}{\lim} \; G_k\left(t\right)=0$.
\begin{itemize}
\item Proof of v)
\end{itemize}
\begin{IEEEeqnarray}{rCl}
\underset{t\rightarrow \infty}{\lim} \; t \, G_k\left(t\right) &=& \underset{t\rightarrow \infty}{\lim} \;  \frac{G_k^{'}\left(t\right)}{\left(\frac{1}{t}\right)^{'}} \label{EA21} \\
                                                          &=& \underset{t\rightarrow \infty}{\lim} \;  \frac{-\frac{1-F_{k}\left(t\right)}{t^{2}}}{-\frac{1}{t^{2}}} \nonumber\\
                                                          &=& 0, \nonumber
\end{IEEEeqnarray}
where we have used l'H\^opital rule to obtain \eqref{EA21}.
\begin{itemize}
\item Proof of vi)
\end{itemize}
\begin{IEEEeqnarray}{rCl}
\underset{t\rightarrow \infty}{\lim} \; \frac{t \, G_k\left(t\right)}{1-F_{k}\left(t\right)} &=& \underset{t\rightarrow \infty}{\lim} \;  \frac{G_k\left(t\right)}{\frac{1-F_{k}\left(t\right)}{t}} \nonumber \\
                                                          &=& \underset{t\rightarrow \infty}{\lim} \;  \frac{-\frac{1-F_{k}\left(t\right)}{t^{2}}}{-\frac{t \, f_{k}\left(t\right)  +1-F_{k}\left(t\right)}{t^{2}}} \label{EA22}\\
                                                          &=& \frac{1}{1+l_k},
\end{IEEEeqnarray}
where we have, again, used l'H\^opital rule to obtain \eqref{EA22}, since $\underset{t \rightarrow \infty}{\lim} \; \frac{t \, f_{k}\left(t\right)}{1-F_{k}\left(t\right)}=l_k$. This completes the proof of Lemma \ref{L3}.
\end{IEEEproof}\

We now go back to the proof of Theorem \ref{T2}. Recall that $\lambda_k\left(\bar{P}_k\right)$ is the minimizer of the dual optimization problem defined by:
\begin{equation}\label{E4B4}
\underset{t \, > \, 0}{\min}\;{\left\{I_k\left(t\right)-t \, \left(G_k\left(t\right)-\bar{P}_k\right)\right\}},
\end{equation}
where $I_k(\cdot)$ is the function defined on $(0,\infty)$ by $I_k\left(t\right)=\int_{t}^{\infty}{\log{\left(\frac{t}{t}\right)f_k(t) \, \mathrm{d}t}}$. For this particular optimization problem, it can be easily verified that
\begin{equation}\label{E4B5}
\frac{\mathrm{d}{I_k}}{\mathrm{d}{t}}=t \; \frac{\mathrm{d}{G_k}}{\mathrm{d}{t}},
\end{equation}
for all $t>0$. Now, for any $t>0$, the following equalities hold true:
\begin{IEEEeqnarray}{rCl}
I_k\left(t\right)&=& \int_{+\infty}^{t}\mathrm{d}I_k\left(u\right) \label{E4B6} \\
							 &=& \int_{+\infty}^{t} u \,  \mathrm{d}G_k\left(u\right) \label{E4B7} \\
                             &=& \int_{0}^{G_k(t)} G_k^{-1}\left(u\right) \,  \mathrm{d}u, \label{E4B71}
\end{IEEEeqnarray}
where $G_k^{-1}(\cdot)$ is the inverse function of $G_k(\cdot)$. Equality \eqref{E4B6} is true because $\underset{t\rightarrow \infty}{\lim} \; I_k\left(t\right)=0$, whereas \eqref{E4B7} follows from \eqref{E4B5}; and \eqref{E4B71} is obtained by change of variables. Applying \eqref{E4B71} to $t=\lambda_k\left(\bar{P}_k\right)$ and using the fact that $I_k\left(\lambda_k\left(\bar{P}_k\right)\right)=C_k\left(\bar{P}_k\right)$, we obtain
\begin{equation}\label{E5A02AB}
C_k\left(\bar{P}_k\right)=\int_{0}^{\bar{P}_k}{G_k^{-1}(t) \, \mathrm{d}t}.
\end{equation}
To prove \eqref{E4B1} in Theorem \ref{T2}, it suffices to we use \eqref{E5A02AB} in order to verify that $\underset{\bar{P}_k\rightarrow 0} {\lim}\frac{C_k\left(\bar{P}_k\right)}{\bar{P}_k \, \lambda_k\left(\bar{P}_k\right)}=1+\frac{1}{l_k}$. This can be shown as follows:
\begin{IEEEeqnarray}{rCl}
\underset{\bar{P}_k\rightarrow 0} {\lim}\frac{C_k\left(\bar{P}_k\right)}{\bar{P}_k \, \lambda_k\left(\bar{P}_k\right)}&=&\underset{\bar{P}_k\rightarrow 0} {\lim}\frac{\int_{0}^{\bar{P}_k}{G_k^{-1}(t) \, \mathrm{d}t}}{\bar{P}_k \, G_k^{-1}\left(\bar{P}_k\right)} \nonumber \\
                                                                                                          &=& \underset{\bar{P}_k\rightarrow 0} {\lim}\frac{G_k^{-1}\left(\bar{P}_k\right)}{G_k^{-1}\left(\bar{P}_k\right) \, \left(1-\frac{\bar{P}_k \, G_k^{-1}\left(\bar{P}_k\right)}{1-F_{k}\left(G_k^{-1}\left(\bar{P}_k\right)\right)}\right)} \label{E4B9} \\
                                                                                                          &=& \frac{1}{1-\frac{1}{1+l_k}}, \label{E4B10}
\end{IEEEeqnarray}
where \eqref{E4B9} follows by the l'H\^opital rule and where \eqref{E4B10} follows from vi) in Lemma \ref{L3}. This establishes \eqref{E4B1} in Theorem \ref{T2}. Next, we prove that the on-off power given by \eqref{E4B12} is capacity-achieving. We treat the cases $l_k \rightarrow \infty$ and $0<l_k<\infty$ separately.

For the class of fading we are interested in, note that if $l_k \rightarrow \infty$, then $\tau_k$ in \eqref{E4B12} is equal to $\lambda_k\left(\bar{P}_k\right)$ and we have:
\begin{IEEEeqnarray}{rCl}
\frac{\bar{P}_k}{1-F_k\left(\tau_k\right)}&=&\frac{\bar{P}_k}{1-F_k\left(\lambda_k\left(\bar{P}_k\right)\right)} \nonumber \\
                                          &=& \frac{G_k\left(\lambda_k\left(\bar{P}_k\right)\right)}{1-F_k\left(\lambda_k\left(\bar{P}_k\right)\right)} \nonumber \\
                                          &\leq& \frac{1}{\lambda_k\left(\bar{P}_k\right)} \label{E4B102},
\end{IEEEeqnarray}
where \eqref{E4B102} follows from the proof of iv in Lemma \ref{L3}. Since $\lambda_k\left(\bar{P}_k\right) \rightarrow \infty$ as $\bar{P}_k \rightarrow 0$, then $\frac{\bar{P}_k}{1-F_k\left(\tau_k\right)}  \rightarrow 0$ as $\bar{P}_k \rightarrow 0$. The rate achieved by the on-off power policy can be computed as follows:
\begin{IEEEeqnarray}{rCl}
R_k &=& \int_{\tau_k}^{\infty}{\log{\left(1+ \gamma_k \; \frac{\bar{P}_k}{1-F_k\left(\tau_k\right)}\right)} \;  f_k(\gamma_k) \; \mathrm{d} \gamma_k } \nonumber \\
             &=& \int_{\lambda_k\left(\bar{P}_k\right)}^{\infty}{\log{\left(1+ \gamma_k \; \frac{\bar{P}_k}{1-F_k\left(\lambda_k\left(\bar{P}_k\right)\right)}\right)} \;  f_k(\gamma_k) \; \mathrm{d} \gamma_k } \nonumber \\
             &\geq&  \int_{\lambda_k\left(\bar{P}_k\right)}^{\infty}{\left(1-\epsilon\right) \; \gamma_k \; \frac{\bar{P}_k}{1-F_k\left(\lambda_k\left(\bar{P}_k\right)\right)}} \; f_k(\gamma_k) \; \mathrm{d} \gamma_k \label{E4B103} \\
             &\geq& \left(1-\epsilon\right) \; \lambda_k\left(\bar{P}_k\right) \; \bar{P}_k, \label{E4B104}
\end{IEEEeqnarray}
where \eqref{E4B103} follows because $\underset{\bar{P}_k \rightarrow 0}{\lim} \, \frac{\bar{P}_k}{1-F_k\left(\tau_k\right)} =0$ and thus $\forall \gamma_k \in [\tau_k,\infty)$, we have $\log{\left(1+\gamma_k \, \frac{\bar{P}_k}{1-F_k\left(\tau_k\right)}\right)} \approx \gamma_k \, \frac{\bar{P}_k}{1-F_k\left(\tau_k\right)}$, or equivalently $\left |\frac{\log{\left(1+\gamma_k \, \frac{\bar{P}_k}{1-F_k\left(\tau_k\right)}\right)}}{\gamma_k \, \frac{\bar{P}_k}{1-F_k\left(\tau_k\right)}}-1 \right \vert \leq \epsilon$, for all $\epsilon >0$, at sufficiently low $\bar{P}_k$. By taking the limits on both sides of \eqref{E4B104} as $\epsilon \rightarrow 0$, we establish that a rate equal to $ \lambda_k\left(\bar{P}_k\right)  \; \bar{P}_k$ is asymptotically achievable. This rate corresponds to the asymptotic capacity described by Theorem \ref{T2} when $l_k \rightarrow \infty$ and hence is the best rate one can achieve.

On the other hand, if $0<l_k<\infty$, then the following statements hold true $\forall t \in [t_0,\infty)$, for some $t_0$ sufficiently large:
\begin{IEEEeqnarray}{rCCCl}
\underset{t\rightarrow \infty}{\lim} \, \frac{t \, f_{k}(t)}{1-F_{k}(t)}=l_k &\Leftrightarrow& \frac{f_k(t)}{1-F_k(t)} &\approx& \frac{l_k}{t} \nonumber \\
&\Leftrightarrow& -\frac{\mathrm{d}}{\mathrm{d} \, t}\left(\log\left(1-F_k(t)\right)\right) &\approx& l_k \, \frac{\mathrm{d}}{\mathrm{d} \, t}\left(\log\left(t\right)\right) \nonumber \\
&\Rightarrow& - \log\left(\frac{1-F_k(t)}{1-F_k(t_0)}\right) &\approx& l_k \, \log\left(\frac{t}{t_0}\right) \label{E4B105} \\
&\Leftrightarrow& - \log\left(1-F_k(t)\right) &\approx& l_k \, \log\left(t\right) \label{E4B106} \\
&\Leftrightarrow& \frac{1}{1-F_k(t)} &\approx& A\, t^{l_k} \label{E4B107}
\end{IEEEeqnarray}
where \eqref{E4B105} follows because if $f(t) \approx g(t)$, then $\forall \epsilon >0$, there exists a certain $t_0$ such that $\forall t \geq t_0$, we have
\begin{equation}\label{E4B110}
\left(1-\epsilon\right) \, g(t) \, < f(t) \, < \left(1+\epsilon\right) \, g(t).
\end{equation}
Integrating \eqref{E4B110} between $t_0$ and $t > t_0$, we establish that
\begin{equation}\label{E4B112}
\int_{t_0}^{t} g(u) \, \mathrm{d}u \approx  \int_{t_0}^{t} f(u) \, \mathrm{d}t.
\end{equation}
The equivalence \eqref{E4B106} is due to the fact that $t_0$ is a constant that vanishes as $t \rightarrow \infty$ whereas $A$ in \eqref{E4B107} is some constant. Therefore, the fixed power in the on-off power policy in this case can be computed as follows:
\begin{IEEEeqnarray}{rCl}
\frac{\bar{P}_k}{1-F_k\left(\tau_k\right)}&=&\frac{\bar{P}_k}{1-F_k\left(\left(1+\frac{1}{l_k}\right)\lambda_k\left(\bar{P}_k\right)\right)} \nonumber \\
                                          &\approx& \frac{\left(1+\frac{1}{l_k}\right)^{l_k} \, \bar{P}_k}{1-F_k\left(\lambda_k\left(\bar{P}_k\right)\right)}  \label{E4B114},
\end{IEEEeqnarray}
where \eqref{E4B114} follows from \eqref{E4B107}. Since $\frac{\bar{P}_k}{1-F_k\left(\lambda_k\left(\bar{P}_k\right)\right)} \rightarrow 0$ as $\bar{P}_k \rightarrow 0$ due to \eqref{E4B102}, then so does $\frac{\bar{P}_k}{1-F_k\left(\tau_k\right)}$. Following similar lines as in the case where $l_k\rightarrow \infty$, it can be shown that a rate equal to
\begin{IEEEeqnarray}{rCl}
R_k  &\geq& \left(1-\epsilon\right) \; \tau_k \; \bar{P}_k  \nonumber \\
              &\approx& \left(1-\epsilon\right) \;  \left(1+\frac{1}{l_k}\right) \; \lambda_k\left(\bar{P}_k\right) \; \bar{P}_k \label{E4B116}
\end{IEEEeqnarray}
is achievable. The RHS of \eqref{E4B116} coincides with the asymptotic capacity in Theorem \ref{T2} as an arbitrary small $\epsilon$ can be chosen. This completes the proof of Theorem \ref{T2}.

\section*{Acknowledgment}
The authors would like to thank the editor Dr. Rui Zhang for volunteering his time to handle this paper and the anonymous reviewers for their valuable comments that have enhanced the technical quality and the lucidity of this paper.

\begin{IEEEbiography}[{\includegraphics[width=1in,height
=1.25in,clip,keepaspectratio]{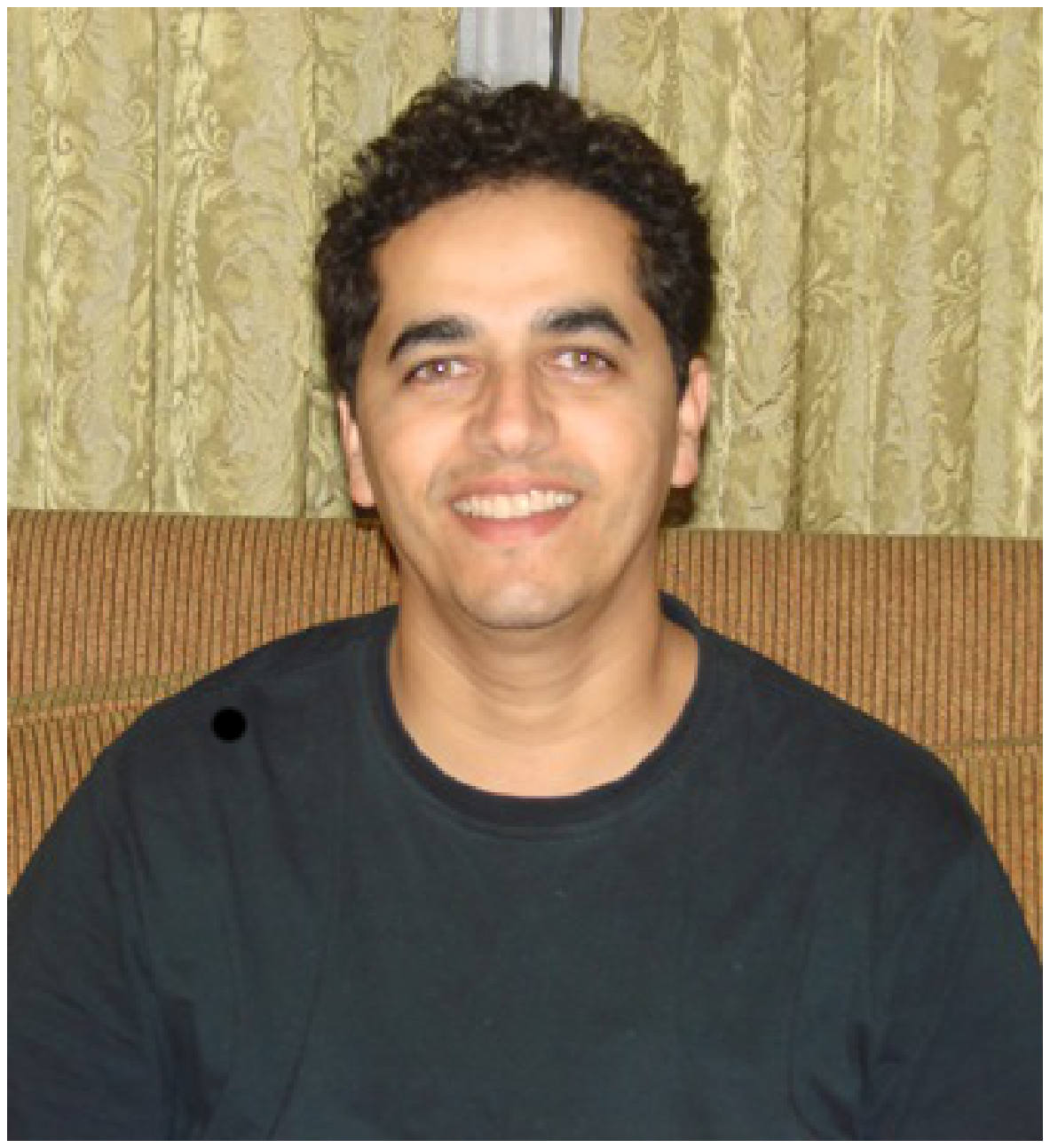}}]{Zouheir Rezki} (S'01, M'08, SM'13) was born in
Casablanca, Morocco. He received the Diplome
d'Ing\'enieur degree from the \'Ecole Nationale de
l'Industrie Min\'erale (ENIM), Rabat, Morocco, in
1994, the M.Eng. degree from  \'Ecole de Technologie
Sup\'erieure, Montreal, Qu\'ebec, Canada, in 2003, and
the Ph.D. degree from  \'Ecole Polytechnique, Montreal,
Qu\'ebec, in 2008, all in electrical engineering.
From October 2008 to September 2009, he was a
postdoctoral research fellow with Data Communications
Group, Department of Electrical and Computer
Engineering, University of British Columbia. He is now a Research Scientist 
at King Abdullah University of Science and Technology (KAUST),
Thuwal, Mekkah Province, Saudi Arabia. His research interests include:
performance limits of communication systems, cognitive and sensor networks,
physical-layer security, and low-complexity detection algorithms.
\end{IEEEbiography}

\begin{IEEEbiography}[{\includegraphics[width=1in,height
=1.25in,clip,keepaspectratio]{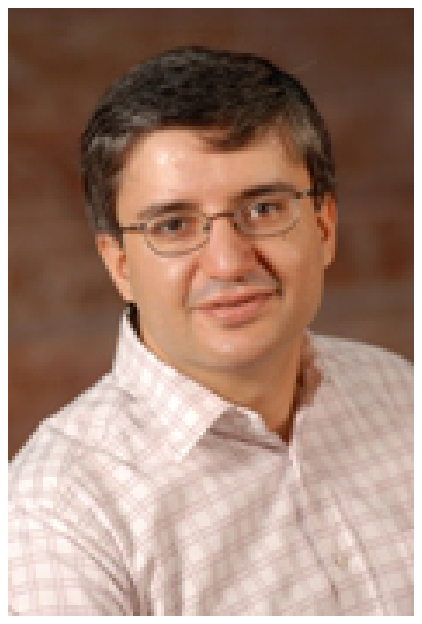}}]{Mohamed-Slim Alouini} (S'94, M'98, SM'03, F'09) was
born in Tunis, Tunisia. He received the Ph.D. degree in Electrical Engineering
from the California Institute of Technology (Caltech), Pasadena,
CA, USA, in 1998. He served as a faculty member in the University of Minnesota,
Minneapolis, MN, USA, then in the Texas A$\&$M University at Qatar,
Education City, Doha, Qatar before joining King Abdullah University of
Science and Technology (KAUST), Thuwal, Makkah Province, Saudi
Arabia as a Professor of Electrical Engineering in 2009. His current
research interests include the modeling, design, and
performance analysis of wireless communication systems.
\end{IEEEbiography}

\end{document}